\documentclass[prd,aps,preprint,showpacs,nofootinbib,superscriptaddress,tightenlines]{revtex4-1}
\usepackage{amssymb,amsmath}
\usepackage{hyperref}
\usepackage{url}
\usepackage{graphicx,subfigure}
\newcommand{\ul}[1]{\underline{#1}}
\newcommand{\avg}[1]{\left< #1 \right>}
\newcommand{\ket}[1]{\left| #1 \right>} 
\newcommand{\braket}[2]{\left< #1 \vphantom{#2} \right| \left.\! #2 \vphantom{#1} \right>}
\newcommand{\bra}[1]{\left< #1 \right|} 
\def\eq#1{{Eq.~(\ref{#1})}}
\def\fig#1{{Fig.~\ref{#1}}}
\def\sec#1{{Sec.~\ref{#1}}}

\begin{document}
\title{\textbf{Radiative $p_{\perp}$-broadening of high-energy quarks and gluons in QCD matter}}
\author{Tseh Liou}
\affiliation{Department of Physics, Columbia University, New York, NY 10027, United States}
\author{A.H.~Mueller}
\affiliation{Department of Physics, Columbia University, New York, NY 10027, United States}
\author{Bin Wu}
\affiliation{Institut de Physique Th\'eorique, CEA Saclay, F-91191 Gif-sur-Yvette, France}

\begin{abstract}
We study radiative $p_{\perp}$-broadening of high-energy quarks passing through hot and cold QCD matter. With $L$ the length of the matter and $l_{0}$ the size of constituents of the matter we find $\avg{p_{\perp}^{2}}$ has both double logarithmic terms, $\ln^{2}(L/l_{0})$, and single logarithmic terms, $\ln(L/l_{0})$, coming from gluon radiation induced by the matter. We use a (slight) extension of a formalism developed by B.~Zakharov for studying energy loss, a formalism which, for much of our calculation, reduces to a simple dipole scattering analysis. We estimate the radiative contribution to be a sizable correction to the nonradiative value of $\avg{p_{\perp}^{2}}$. We also carry out a resummation of the double logarithmic terms that we find, and we briefly discuss running coupling effects which appear here in a rather unusual way.
\end{abstract}

\maketitle

\section{\label{sec:intro} Introduction}

High-energy quarks and gluons passing through hot and cold nuclear matter have their transverse momentum distributions broadened due to multiple, inelastic, scattering with the constituents of the nuclear matter \cite{Bodwin,bdmpspt}. This transverse momentum broadening can be nicely expressed in terms of the elastic scattering of a quark or gluon dipole in the nuclear matter as given in \eq{eq:dN} below.

In fact the relationship between, say, quark transverse momentum broadening and the elastic scattering of a quark-antiquark dipole given in \eq{eq:dN} appears to be a rather general relationship in gauge theories \cite{Kovchegov,Mueller}. In this paper we will exploit this relationship to evaluate $p_{\perp}$-broadening where the broadening may come from gluon radiation as well as from multiple scattering. More precisely, $p_{\perp}$-broadening from inelastic multiple scatterings is given as $p_{\perp}^{2}=\hat{q}L$ where $L$ is the length of the material and $\hat{q}$ is the transport coefficient corresponding to the hard parton, quark or gluon, passing through the medium, hot or cold QCD matter. In this paper we evaluate next to leading order corrections which are enhanced by single and double logarithms, that is radiative corrections of size $(\alpha_{s} N_{c}\hat{q}L/\pi)\ln^{2}(L/l_{0})$ and $(\alpha_{s} N_{c}\hat{q}L/\pi)\ln (L/l_{0})$ where $l_{0}$ is the ``size'' of the scatterers in the medium. The full expression of the radiative correction to the $p_{\perp}$-broadening is shown in \eq{eq:psd}. Although $l_{0}$ serves as an infrared cut-off in the calculation, we shall see explicitly that the double and single logarithmic terms are independent of the exact definition of $l_{0}$. 

To a large extent this paper is motivated by an earlier paper by one of us (BW) who dealt with the same problem in Ref.~\cite{BW}. That paper formulated the next to leading order $p_{\perp}$-broadening problem in terms of the BDMPS formalism \cite{bdmpspt,bdmpsenergy,bdmpsz} developed for calculating radiative energy loss of high-energy partons passing through nuclear matter. However, the BDMPS formalism has a shortcoming, namely the single scattering term with the medium is absent. For radiative energy loss this is not a difficulty, at least for a sufficiently large medium, since the dominant contribution comes from multiple scattering. But in the present context the single scattering term is crucial. There is a second formalism, developed by B.~G.~Zakharov \cite{zlpm,zrad} for calculating radiative energy loss. While the formalism of Zakharov, in contrast to BDMPS, does not \textit{explicitly} eliminate the single scattering term the final answers for radiative energy loss are identical in the two formalisms in the harmonic oscillator approximation. However, for radiative $p_{\perp}$-broadening this is not the case and as a result the Zakharov formalism seems better suited to the broadening problem. We could get our result \eq{eq:psd} by taking the calculation of Ref.~\cite{BW} and adding the single scattering term missing in Ref.~\cite{BW}. We have done this but we believe our present discussion to be simpler.

The Zakharov formalism is a dipole-like formalism with the difference being that the quark-antiquark-gluon system, dominant in radiative energy loss or radiative $p_{\perp}$-broadening, is not frozen in impact parameter over the times of the scatterings as in the case in normal high-energy scattering uses of the dipole formalism \cite{dipolebfkl,book}. However, in the single scattering approximation in our current application a Zakharov type formalism is the same as the normal dipole formalism. As we shall see below all the double logarithmic terms and some of the single logarithmic terms come from single scattering and must be given by a simple dipole evaluation.

It is not hard to see why the single scattering plays such a prominent role in the present calculation. In the dipole picture nonradiative $p_{\perp}$-broadening is given as the Fourier transform of the elastic scattering of a dipole of transverse size $x_{\perp}$ with the medium as given in \eq{eq:dN}. The dipole size which dominates the integral in \eq{eq:dN} is $x_{\perp}\sim 1/Q_{s}$ where $Q_{s}^{2}=\hat{q}L$ so that in terms of the dipole formalism only a few elastic scatterings, over the length $L$ of the medium, are used to determine the typical transverse momentum given by the \textit{many} inelastic scatterings of the quark passing through the medium. If we now add a radiative gluon to the dipole the gluon can be characterized by its lifetime and the transverse size ($B_{\perp}$) that it makes with the original quark-antiquark dipole of size $x_{\perp}$. In order to get a double logarithmic contribution it is necessary to have logarithmic integrals over both the lifetime of the gluon fluctuation and $B_{\perp}$. It is clear that $B_{\perp}$ must be much larger than $x_{\perp}$, because the quark-antiquark dipole only has, typically, a single scattering over a time $L$ and now we wish to have the quark-antiquark-gluon system to interact with the medium over a time, or distance, short compared to $L$. However it is known that when $B_{\perp}\gg x_{\perp}$ the probability of a quark-antiquark dipole to emit a gluon goes as $x_{\perp}^{2}dB^{2}_{\perp}/B^{4}_{\perp}$. This is not a logarithmic integral, however, it becomes logarithmic if and only if one includes a single scattering, which gives an additional factor of $B^{2}_{\perp}$. However, if we include more than one scattering, the logarithmic structure is destroyed. As we will see in the following calculation there is no double logarithmic contribution from the multiple scattering region.

We are now in a position to indicate which kinematic regions give single and double logarithmic contributions in terms of the regions shown in \fig{fig:tregion} where, for technical convenience, we have chosen to characterize the radiative gluon by its lifetime, $t$, and its energy, $\omega$. The reasons that we divide the phase space into different regions, separated by the boundaries $(a)$, $(b)$ and $(c)$, will be explained in detail in \sec{sec:r}. As indicated in the figure the lifetime can be between zero and $L$ while the energy of the gluon is between $0$ and $\hat{q}L^{2}$. The interior of the region between lines $(a)$, $(b)$ and $(c)$ gives the double logarithmic contribution to the radiative contribution to $\avg{p_{\perp}^{2}}$ of the quark. Single logarithmic contribution comes from regions close to the lines $(a)$, $(b)$ and $(c)$. The line $(b)$ is the transition between single scattering dominance, below $(b)$, and multiple elastic scattering where the $S$ in \eq{eq:dN} becomes small, above $(b)$. The region close to line $(a)$ is a single scattering region where the transverse distance between the gluon and either the quark or antiquark is comparable to the quark-antiquark separation $x_{\perp}$. The region around line $(c)$ is the region where the lifetime of the gluon is on the order of the proton's size, for cold matter. In this region the transport coefficient can no longer be treated as a constant since the gluon distribution included in $\hat{q}$ is evaluated at large $x$-values. This region is, however, dominated by single scattering with the medium.

We should comment on some of the limitations and subtleties of our calculation. \textbf{(a)} We have done our calculation in a context of taking the transport coefficient $\hat{q}$ to be scale independent. This is not necessarily the best choice, so in \sec{sec:running} we do a running coupling calculation of the double logarithmic radiative terms. Curiously the double logarithmic contribution to $\avg{p_{\perp}^{2}}$ turns out to be exactly a factor of two greater in the running coupling case as compared to fixed coupling and, remarkably, no $\ln\ln$ terms appear. We do not understand how to do single logarithmic terms in the present running coupling context. \textbf{(b)} In all the regions in which there are important contributions it is only near the boundary (a) in \fig{fig:tregion} that the emitted gluon transverse coordinate gets as close to either the quark or antiquark as the size of the quark-antiquark separation. Thus except for the boundary $(a)$ calculation given in \sec{sec:boundarya} we can define a mean $p_{\perp}^{2}$, $\avg{p_{\perp}^{2}}$, using \eq{eq:avgp}. However near the boundary $(a)$ there is an $x_{\perp}^{2}\ln x_{\perp}^{2}$ dependence for $S(x_{\perp})$ so that the mean of $p_{\perp}^{2}$ no longer makes sense. This corresponds to a $p_{\perp}^{-4}$ behavior for large $p_{\perp}^{2}$ in $d N/d^{2}p_{\perp}$. An almost identical issue has been discussed earlier in Refs.~\cite{bdmpspt,Arnold} where it was suggested to replace the idea of a mean $p_{\perp}^{2}$ by that of a typical $p_{\perp}^{2}$. This amounts to taking the $\nabla_{x_{\perp}}^{2}$ in \eq{eq:avgp} on all terms other than the $\ln x_{\perp}^{2}$ and at the end setting $x_{\perp}^{2}$ to its normal scale, in our case $x_{\perp}^{2}\simeq 4/Q_{s}^{2}$. In the calculation given in \sec{sec:boundarya} we follow this prescription and leave the answer, in \eq{eq:boundarya} and \eq{eq:psd}, in terms of $x_{\perp}^{2}$. In our final estimate we take $x_{\perp}^{2}=4/Q_{s}^{2}$ to get \eq{eq:est}. \textbf{(c)} The double logarithmic terms can be resummed, and this is done in \sec{sec:resum}. However, it seems that one needs media lengths rather large for this resummation to be useful. For lengths of about $5\, \mathrm{fm}$ the first leading double logarithmic term gives the dominant contribution. \textbf{(d)} The radiative corrections are reasonably important, comparable to the values of $p_{\perp}^{2}$ coming from nonradiative multiple scattering. This is similar to the conclusion of Ref.~\cite{BW}.

An outline of the paper is as follows: In \sec{sec:ds} we briefly describe $p_{\perp}$-broadening of a quark due to multiple inelastic scatterings and remind the reader that these inelastic scatterings can be recast in terms of the elastic scattering of a dipole in the medium. In \sec{sec:g} we give, but do not derive, the general formalism which we use, a generalization of the formalism that Zakharov developed for radiative energy loss. In \sec{sec:r} we quantitatively delineate the different kinematic regions for the radiated gluon. In \sec{sec:d} we evaluate the double logarithmic contribution to the average transverse momentum broadening, $\avg{p_{\perp}^{2}}$, of the quark due to gluon radiation. In \sec{sec:s} we evaluate the single logarithmic contributions to $\avg{p_{\perp}^{2}}$. In \sec{sec:f} we give the final result and do a rough estimate of the radiative correction. In \sec{sec:re} we discuss resummation and running coupling effects. In \sec{sec:der} we give a derivation of our formalism and the relationship to that of Ref.~\cite{BW}.

\section{\label{sec:ds}$p_{\perp}$-broadening and dipole scattering}

\subsection{A high-energy quark entering a nucleus from outside}
We begin by considering a high-energy quark passing through a large nucleus and scattering either inelastically, one gluon exchange, or elastically, two gluon exchange, with the nucleons in the nucleus. For the moment we work in the McLerran-Venugopalan (MV) \cite{mv} approximation where there is no small-$x$ evolution. If the quark enters the nucleus with no transverse momentum and exits having transverse momentum $p_{\perp}$, then the distribution of transverse momenta is given by
\begin{equation}
  \frac{dN}{d^{2}p_{\perp}}=\int\frac{d^{2}x_{\perp}}{(2\pi)^{2}}\, e^{-ip_{\perp}\cdot x_{\perp}}S(x_{\perp})
  \label{eq:dN}
\end{equation}
where $S(x_{\perp})$ is the $S$-matrix for elastic scattering of a quark-antiquark dipole of transverse size $x_{\perp}$ on the nucleus. In the MV model the nucleons in the nucleus are uncorrelated so that
\begin{equation}
  S(x_{\perp})=\exp\big[S_{1}(x_{\perp})\big]
  \label{eq:snucleus}
\end{equation}
where $S_{1}(x_{\perp})$ is the elastic dipole-nucleon scattering $S$-matrix which can be evaluated to be
\begin{equation}
  S_{1}(x_{\perp})=-\frac{x_{\perp}^{2}Q_{s}^{2}}{4}
  \label{eq:snucleon}
\end{equation}
with the quark saturation momentum given by \cite{bdmpspt}
\begin{equation}
  Q_{s}^{2}=\frac{4\pi^{2}\alpha_{s} C_{F}}{N_{c}^{2}-1}L\rho xG(x,1/x_{\perp}^{2}).
  \label{eq:sat}
\end{equation}

In \eq{eq:sat} $L$ is the length of nuclear matter that the quark passes through, $\rho$ the nucleon density, and $G(x,1/x_{\perp}^{2})$ the gluon distribution of the nucleon. While \eq{eq:dN} is not a general formula it was shown in Refs.~\cite{Kovchegov,Mueller} that it is correct in leading and next to leading orders of small-$x$ evolution, thus carrying its validity well beyond the MV model.

\subsection{A high-energy quark produced in a nucleus}

The problem of $p_{\perp}$-broadening of a quark which is produced in a nucleus, say in a proton-nucleus or a nucleus-nucleus collisions, is quite different from that of a quark entering the nucleus from outside. For a high-energy quark entering the nucleus from outside, small-$x$ evolution corresponds to an evolved wave function of the corresponding dipole as it enters the nucleus. That is, small-$x$ evolution mainly occurs before the quark, or corresponding dipole, reaches the target. However, a quark produced in the nucleus through a hard collision is produced as a bare quantum and small-$x$ evolution is much more limited, and that evolution occurs in the nuclear environment. It is precisely these (small-$x$) radiative corrections to $p_{\perp}$-broadening of a (bare) quark produced in a hard collision which is the topic of this paper. At the MV level there is no distinction between a quark entering a nucleus from outside and a quark created in the nucleus so that Eqs.~(\ref{eq:dN}), (\ref{eq:snucleus}), (\ref{eq:snucleon}) and (\ref{eq:sat}) are valid while \eq{eq:dN} should remain valid even in the presence of radiation in the nuclear medium, that is beyond the MV approximation.

\section{\label{sec:g}$p_{\perp}$-broadening due to gluon emission: the general formalism}

We have seen that quark $p_{\perp}$-broadening due to multiple scattering of the quark in a nuclear medium can be succinctly expressed in terms of the elastic scattering of a quark-antiquark dipole in the medium. The quark part of the dipole corresponding to the quark in the original amplitude and the antiquark in the complex conjugate amplitude. This same picture holds true when radiative corrections are included \cite{Kovchegov,Mueller}. $p_{\perp}$-broadening due to multiple scattering and single gluon emission is given by elastic scattering of a dipole in the medium where the dipole emits, and later absorbs, a gluon as it passes through the medium. We shall focus on soft gluon emission in which case the quark and antiquark making up the dipole can be viewed as fixed in transverse coordinate space during the emission and absorption of the gluon. The gluon, however, moves in transverse coordinate space during its lifetime, starting from either the quark or antiquark and ending again on one of the two parts of the quark-antiquark dipole. During the time that the gluon fluctuation lives, the quark-antiquark-gluon system elastically scatters off nucleons in the nucleus. While this three body scattering can be done exactly, in this section we shall limit our discussion to the large $N_{c}$ and soft gluon limits where the picture is especially simple. A more general discussion is given in Appendix~\ref{app:zakharov}.

\begin{figure}[h]
  \centering
  \includegraphics[width=6cm]{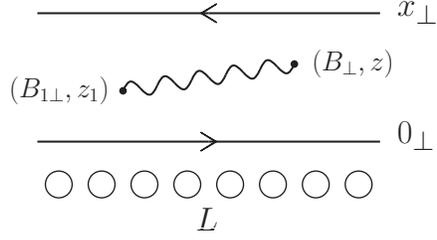}
  \caption{A quark-antiquark-gluon system moves in a nuclear medium with length $L$. The transverse coordinates of the quark and antiquark are $0$ and $x_{\perp}$ respectively. A gluon propagates from $(B_{1\perp},z_{1})$ to $(B_{\perp},z)$, and its motion is described by the Green function $G(B_{\perp},z;B_{1\perp},z_{1})$.}
  \label{fig:singledipole}
\end{figure}
The essential part of the calculation involves the propagation of the quark-antiquark-gluon system through the nuclear medium. We choose the position of the quark to be at the origin in transverse coordinate space and the antiquark to be at $x_{\perp}$ and these positions are fixed during the evolution. Let $G(B_{\perp},z;B_{1\perp},z_{1})$ be the Green function for the propagation of the system from longitudinal position $z_{1}$ to $z$ where the gluon has transverse position $B_{1\perp}$ at $z_{1}$ and transverse position $B_{\perp}$ at $z$. (Since our problem is of relativistic motion longitudinal coordinates and times can be identified.) We normalize $G$ by
\begin{equation}
  G(B_{\perp},z_{1};B_{1\perp},z_{1})=\delta^{(2)}(B_{\perp}-B_{1\perp}).
  \label{eq:Gn}
\end{equation}

$G$ is pictured in \fig{fig:singledipole}. In the large $N_{c}$ limit the evolution of $G$ is straightforward and is given by Ref.~\cite{BW}, with $\omega$ the energy of the gluon,
\begin{equation}
  \frac{\partial}{\partial z}G(B_{\perp},z;B_{1\perp},z_{1})=\bigg[-\frac{i}{2\omega}\nabla_{B_{\perp}}^{2}-\frac{\hat{q}}{4}\big(B^{2}_{\perp}+(B_{\perp}-x_{\perp})^{2}\big)\bigg]G(B_{\perp},z;B_{1\perp},z_{1})
  \label{eq:G}
\end{equation}
where $\hat{q}$ is the quark transport coefficient. $\hat{q}$ can be obtained from \eq{eq:sat} as $\hat{q}=Q_{s}^{2}/L$ and the relationship to the gluon transport coefficient is $\hat{q}_{\mathrm{gluon}}=\hat{q}N_{c}/C_{F}$. The first term on the right hand side of \eq{eq:G} represents free gluon propagation while the second and third terms correspond to scattering of the quark-(antiquark part of the gluon) dipole and the (quark part of the gluon)-antiquark dipole in the medium. It is convenient to rewrite \eq{eq:G} as 
\begin{equation}
  \frac{\partial G}{\partial z}=-\frac{i}{2\omega}\nabla_{B_{\perp}}^{2}G-\frac{\hat{q}}{2}\bigg[\Big(B_{\perp}-\frac{x_{\perp}}{2}\Big)^{2}+\frac{x^{2}_{\perp}}{4}\bigg]G.
  \label{eq:Gc}
\end{equation}

\eq{eq:Gc} is the equation for a two-dimensional harmonic oscillator with an imaginary potential. The solution, with initial condition \eq{eq:Gn}, is
\begin{align}
  G(B_{\perp},z;B_{1\perp},z_{1})&=\frac{i\omega\omega_{0}}{2\pi\sin\omega_{0}t}\exp\Bigg\{\frac{-i\omega\omega_{0}}{2\sin\omega_{0}t}\bigg[\bigg(\Big(B_{\perp}-\frac{x_{\perp}}{2}\Big)^{2}+\Big(B_{1\perp}-\frac{x_{\perp}}{2}\Big)^{2}\bigg)\cos\omega_{0}t\nonumber\\
  &\phantom{==========}-2\Big(B_{\perp}-\frac{x_{\perp}}{2}\Big)\cdot\Big(B_{1\perp}-\frac{x_{\perp}}{2}\Big)\bigg]-\frac{\hat{q}x^{2}_{\perp}t}{8}\Bigg\}
  \label{eq:green}
\end{align}
where 
\begin{equation}
  t=z-z_{1}
\end{equation}
and 
\begin{equation}
  \omega_{0}=\frac{1+i}{\sqrt{2}}\sqrt{\frac{\hat{q}}{\omega}}.
  \label{eq:omega}
\end{equation}

In the large $N_{c}$ limit the $\omega_{0}$ in \eq{eq:omega} is exactly as in Ref.~\cite{BW}. One gets a formula for the radiative contribution to $S(x_{\perp})$, in \eq{eq:dN}, by connecting the gluon at $z_{1}$, and $z$ to the quark and antiquark and by using the usual MV formula during the times when the gluon is not present. Suppose the nuclear matter is located in the longitudinal region $0<z<L$. Then with
\begin{equation}
  S(x_{\perp})=\int \frac{d\omega}{\omega}\, N(x_{\perp},\omega),
  \label{eq:Sx}
\end{equation}
$N(x_{\perp},\omega)$ is given by
\begin{align}
  N(x_{\perp},\omega)&=-\frac{\alpha_{s} N_{c}}{2\omega^{2}}\mathrm{Re}\int^{L}_{0}dz_{2}\, \int^{z_{2}}_{0}dz_{1}\,\nabla_{B_{1\perp}}\cdot \nabla_{B_{2}\perp}\Big[e^{-\hat{q}x_{\perp}^{2}(L-z_{2})/4-\hat{q}x_{\perp}^{2}z_{1}/4}G(B_{2\perp},z_{2};B_{1\perp},z_{1})\nonumber\\
  &\phantom{=====}-G_{0}(B_{2\perp},z_{2};B_{1\perp},z_{1})\Big]\bigg|^{B_{2\perp}=x_{\perp}}_{B_{2\perp}=0}\bigg|^{B_{1\perp}=x_{\perp}}_{B_{1\perp}=0}
  \label{eq:Nx}
\end{align}
where $G_{0}$ is obtained from $G$ by setting $\omega_{0}=\hat{q}=0$ in \eq{eq:green}. A few comments on \eq{eq:Nx}: \textbf{(i)} The overall normalization will be fixed in the next section when we evaluate the single scattering term in $G$ and connect the result with that from the standard dipole scattering formalism. \textbf{(ii)} The contribution in \eq{eq:Nx} will give the radiative contribution to $p_{\perp}$-broadening. One should add this contribution to the purely multiple scattering contribution discussed in \sec{sec:ds}. \textbf{(iii)} The subtraction of the $G_{0}$ term in \eq{eq:Nx} subtracts the medium independent contribution. \textbf{(iv)} \eq{eq:Nx} is a generalization of an analogous formula introduced by Zakharov \cite{zlpm,zrad} to evaluate quark energy loss due to gluon radiation in a nuclear or thermal QCD medium. \textbf{(v)} Note that in the above Zakharov formalism the longitudinal positions (times) $z_{1}$ and $z_{2}$ are the times when the gluon is emitted and absorbed from the quark or antiquark while similar time variables in a BDMPS formalism correspond to the positions of scatterers in the amplitude and complex conjugate amplitude for gluon emission. Thus the two $z$-integrals in \eq{eq:Nx} do not require more than a single scattering, while in a BDMPS formula two or more scatterings are required to get two independent time integrals. \textbf{(vi)} Finally, a more complete derivation of \eq{eq:Nx} is given in \sec{sec:der} and Appendix~\ref{app:zakharov}.

\section{\label{sec:r}The various kinematic regions}

Before going on to evaluate the single and double logarithmic contributions to \eq{eq:Sx} it will be helpful to qualitatively determine the kinematic regions in $\omega$ and $t=z_{2}-z_{1}$ which are important in \eq{eq:Sx} and \eq{eq:Nx}. Clearly $0<t<L$, however when $t\lesssim l_{0}$ we shall see that special care must be taken so that a straightforward use of \eq{eq:Sx} only occurs in $l_{0}<t<L$. At $z_{1}$ and $z_{2}$, $B_{1\perp}$ and $B_{2\perp}$ are taken to be either $0$ or $x_{\perp}$. However, at $z$-values between $z_{1}$ and $z_{2}$ one expects that $B_{\perp}$ may be much larger than $x_{\perp}$, which we always suppose to be of size $x_{\perp}\sim 1/Q_{s}$. We may estimate the typical size of $B_{\perp}$ using free gluon diffusion between $z_{1}$ and $z_{2}$. This gives
\begin{equation}
  B_{\perp}^{2}\sim \frac{t}{\omega}.
  \label{eq:Bperp}
\end{equation}

However, when $B^{2}_{\perp}$ gets too large, $G$, given by \eq{eq:green}, will go to zero exponentially in $B^{2}_{\perp}$. It is not hard to see that $G$ will be small unless
\begin{equation}
  |\omega_{0} t|\lesssim 1\qquad \mathrm{or}\qquad  t\lesssim\sqrt{\frac{\omega}{\hat{q}}}
  \label{eq:tone}
\end{equation}
which can be rewritten in terms of transverse coordinate $B_{\perp}$ as
\begin{equation}
  \label{eq:tonet}
  \omega B_{\perp}^{2}\hat{q}\lesssim \frac{1}{B_{\perp}^{2}}.
\end{equation}
Since $\omega B_{\perp}^{2}\hat{q}$ is the transverse momentum picked up by the gluon over its lifetime. The condition \eq{eq:tonet} simply states that the momentum transferred from the medium to the radiated gluon is less than that carried by the gluon and thus single scattering dominates. This requirement will become clearer when we calculate the $p_{T}$-broadening in the dipole language in \sec{sec:d}. Finally when $B^{2}_{\perp}$ becomes as small as $x^{2}_{\perp}$ the gluon will become very coherent with a part or all of the quark-antiquark dipole and, again, special care will be required. In terms of $t$ and $\omega$ the region $B^{2}_{\perp}\sim x^{2}_{\perp}$ corresponds to, using \eq{eq:Bperp} above and $x^{2}_{\perp}\sim 1/Q_{s}^{2}\sim 1/\hat{q}L$,
\begin{equation}
  t\sim \frac{\omega}{\hat{q}L}.
  \label{eq:ttwo}
\end{equation}

\begin{figure}[h]
  \centering
  \includegraphics[width=9cm]{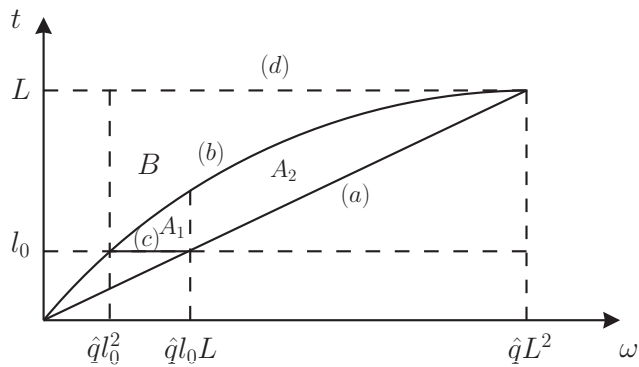}
  \caption{The phase space of the radiated gluon. $t$ and $\omega$ are the lifetime and energy of the gluon, respectively. The double logarithmic region is enclosed by the lines $(a)$, $(b)$ and $(c)$. Regions $A_{1}$ and $A_{2}$ combined together are the region $A$.}
  \label{fig:tregion}
\end{figure}
The various boundaries and regions are shown in \fig{fig:tregion}. The line $(a)$ in that figure is given by $t=\omega/\hat{q}L$ as in \eq{eq:ttwo}. Line $(b)$ is given by $t=\sqrt{\omega/\hat{q}}$ as in \eq{eq:tone} while line $(c)$ given by $t=l_{0}$ is given by the size of the nucleons in cold matter or by the inverse of the temperature in hot matter.

As we shall shortly see the region $A$, enclosed by the boundary lines $(a)$, $(b)$ and $(c)$, is a region where \eq{eq:Nx} gives a double logarithmic contribution to \eq{eq:Sx}. This double logarithmic region is dominated by the single scattering contribution to $N(x_{\perp},\omega)$. The region $B$, above line $(b)$, is a region of strong absorption of the quark-antiquark-gluon system due to the separation of the gluon from the quark-antiquark pair being very large. In the next section we shall first evaluate \eq{eq:Sx} in the region $A$, a relatively easy task, to obtain the double logarithmic contribution. Then we shall, in turn, integrate \eq{eq:Nx} across the boundary lines $(a)$, $(b)$ and $(c)$ which, after doing the remaining logarithmic integration, will give the complete single logarithmic contributions. The constant, non-logarithmic contributions coming near the intersections of the lines $[(a),(b)]$, $[(a),(c)]$ and $[(b),(c)]$ are beyond what we are currently able to do.

\section{\label{sec:d}The double logarithmic contribution to $\avg{p^{2}_{\perp}}$}

In this section we shall evaluate the leading radiative contribution to
\begin{equation}
  \avg{p_{\perp}^{2}}=\int d^{2}p_{\perp}\, p_{\perp}^{2}\frac{dN}{d^{2}p}.
  \label{eq:pdef}
\end{equation}
Using \eq{eq:dN} it is straightforward to get
\begin{equation}
  \avg{p_{\perp}^{2}}=-\nabla_{x_{\perp}}^{2}S(x_{\perp})\bigg|_{x_{\perp}=0}
  \label{eq:avgp}
\end{equation}
which we shall now evaluate using \eq{eq:Sx} and \eq{eq:Nx}. In the double logarithmic region, region $A$ of \fig{fig:tregion}, the typical values of $B_{\perp}$ are much larger than $x_{\perp}$ so that we evaluate the differences in $B_{1\perp}$ and $B_{2\perp}$, taken at $0$ and $x_{\perp}$, by expanding $G$ so that \eq{eq:Nx} becomes
\begin{align}
  N(x_{\perp},\omega)&=-\frac{\alpha_{s} N_{c} x_{\perp}^{2}}{4\omega^{2}}\mathrm{Re}\int^{L}_{0}dt\, (L-t)\big(\nabla_{B_{1\perp}}\cdot \nabla_{B_{2\perp}}\big)^{2}\Big[G(B_{2\perp},t;B_{1\perp},0)\nonumber\\
  &\phantom{====}-G_{0}(B_{2\perp},t;B_{1\perp},0)\Big]\bigg|_{B_{1\perp}=B_{2\perp}=0}.
  \label{eq:Nw}
\end{align}
We have dropped the $x_{\perp}^{2}$ term in the exponent since the $x_{\perp}^{2}$ factor in \eq{eq:Nw} is what is necessary to evaluate \eq{eq:avgp}. It is straightforward to evaluate the derivatives of $B_{1\perp}$ and $B_{2\perp}$ in \eq{eq:Nw} as
\begin{align}
  &\big(\nabla_{B_{1\perp}}\cdot \nabla_{B_{2\perp}}\big)^{2}\Big(G(B_{2\perp},t;B_{1\perp},0)-G_{0}(B_{2\perp},t;B_{1\perp},0)\Big)\bigg|_{B_{1\perp}=B_{2\perp}=0}\nonumber\\
  &\phantom{=====}=-\frac{i}{\pi}\bigg(\frac{\omega}{t}\bigg)^{3}\Bigg\{\bigg(\frac{\omega_{0}t}{\sin\omega_{0}t}\bigg)^{3}\big[4-\sin^{2}\omega_{0}t\big]-4\Bigg\}
  \label{eq:nG}
  \end{align}
where the final term in \eq{eq:nG} comes from $G_{0}$. Now taking $\omega_{0}t$ small one gets
\begin{equation}
  \label{eq:nGs}
  \big(\nabla_{B_{1\perp}}\cdot \nabla_{B_{2\perp}}\big)^{2}\big(G-G_{0}\big)\bigg|_{B_{1\perp}=B_{2\perp}=0}\simeq \frac{\hat{q}\omega^{2}}{\pi t}.
\end{equation}
Using \eq{eq:nGs} in \eq{eq:Nw} along with \eq{eq:Sx} and \eq{eq:avgp}
\begin{equation}
  \label{eq:pp}
  \avg{p_{\perp}^{2}}=\frac{\alpha_{s} N_{c}}{\pi}\hat{q}L\int^{L}_{l_{0}}\frac{dt}{t}\int^{\hat{q}Lt}_{\hat{q}t^{2}}\frac{d\omega}{\omega}=\frac{\alpha_{s} N_{c}}{8\pi}\hat{q}L\ln^{2}\bigg(\frac{L}{l_{0}}\bigg)^{2}
\end{equation}
where the limits of integration in \eq{eq:pp} are given by the boundaries of the region $A$ in \fig{fig:tregion}. The result in \eq{eq:pp} does not assume the large $N_{c}$ limit. It may be useful to recall what the $\omega$ limits in \eq{eq:pp} correspond to in physical terms. The lower limit $\omega>\hat{q}t^{2}$ can be viewed as $\omega/t>\hat{q}t$. But from diffusion of the gluon this becomes, using \eq{eq:Bperp},
\begin{equation}
  \label{eq:bbpp}
  B_{\perp}^{2}<\frac{1}{\hat{q}t}
\end{equation}
where $B_{\perp}^{2}$ is the maximum transverse distance of the gluon from the quark or antiquark. \eq{eq:bbpp} simply expresses the requirement that $B_{\perp}^{2}$ be small enough that the single scattering approximation should be valid. This is as previously discussed just below \eq{eq:tonet}. Similarly, the upper limit on the $\omega$-integration requires $\omega<\hat{q}Lt$ or, again using \eq{eq:Bperp},
\begin{equation}
  \label{eq:Bx}
  B_{\perp}^{2}>\frac{1}{Q_{s}^{2}}\approx x_{\perp}^{2}.
\end{equation}
\eq{eq:Bx} is the requirement that $B_{\perp}^{2}$ be greater than the transverse size of the quark-antiquark dipole. An accurate, single logarithmic, evaluation of the region $B_{\perp}^{2}\sim x_{\perp}^{2}$ is given in \sec{sec:boundarya}.

Finally, let us see that the answer \ref{eq:pp} can be obtained almost effortlessly using the standard dipole picture for $S(x_{\perp})$. We first note that it is natural to take $x_{\perp}\lesssim 1/Q_{s}$ otherwise $S(x_{\perp})$ will be very small just from the elastic quark-antiquark dipole scattering in the medium. When $x_{\perp}\sim 1/Q_{s}$ there is typically one quark-antiquark dipole scattering with the whole medium. Now in order to get a double logarithmic term from radiation one of the logarithmic integration must correspond to the various lifetimes of the gluon fluctuation. This means that the gluon lifetime is much less than the length, $L$, of the medium. Thus for there to be a significant chance for a scattering of the quark-antiquark-gluon system the gluon separation from the quark and antiquark must be very large to enhance the scattering. In this circumstance the single scattering, but only the single scattering contribution, can give a $dB_{\perp}^{2}/B_{\perp}^{2}\sim dt/t$ logarithmic contribution. (Higher multiple scatterings would give extra powers of $B_{\perp}^{2}$ thus destroy the logarithmic integration.)

\begin{figure}[h]
  \centering
  \includegraphics[width=5cm]{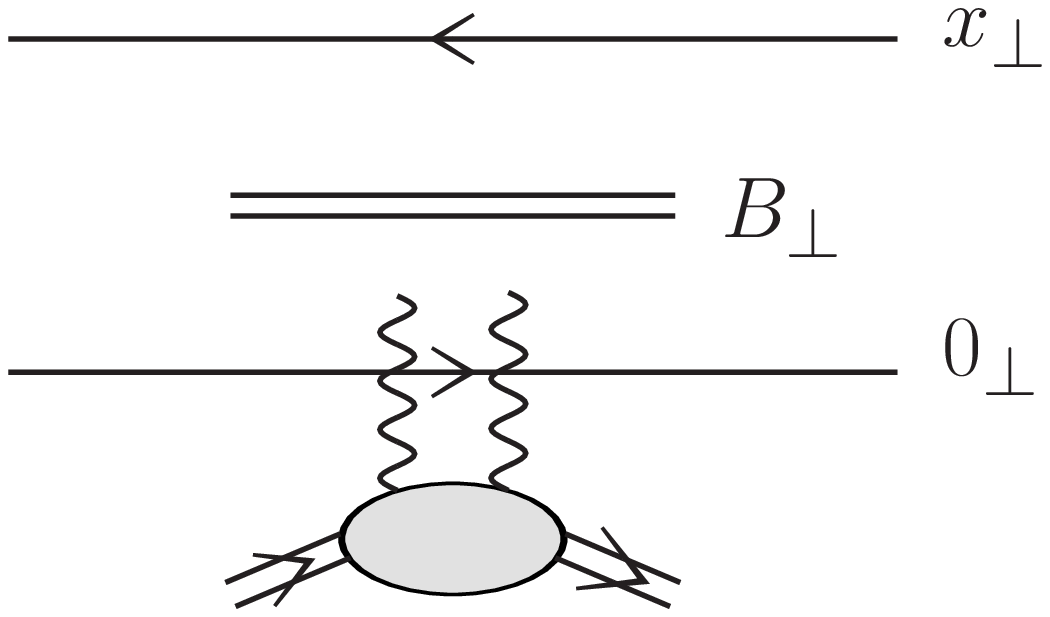}
  \hspace{0.5in}
  \includegraphics[width=5cm]{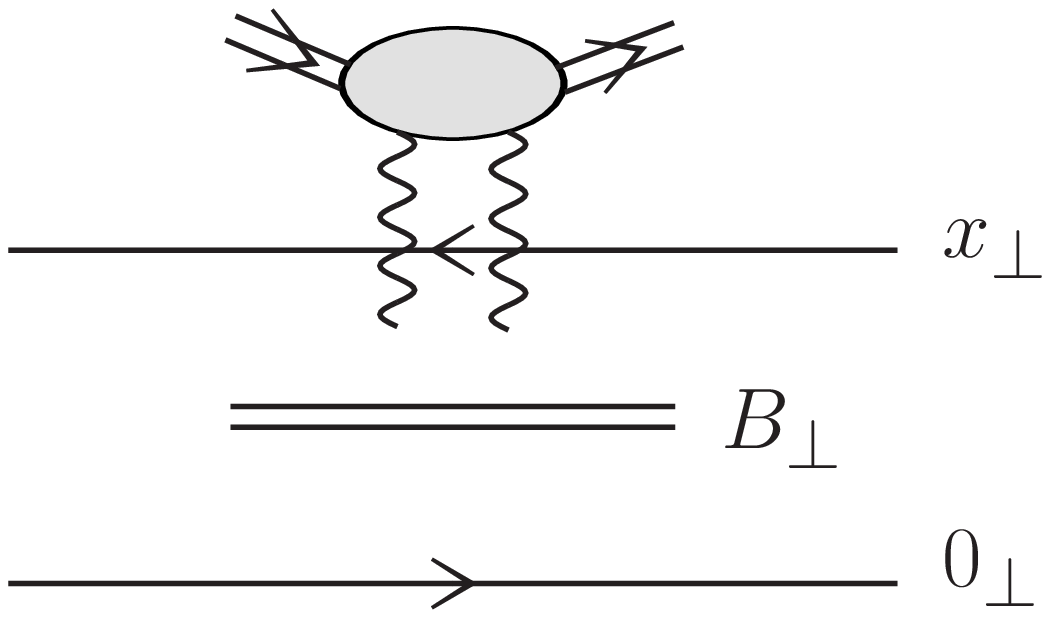}
  \caption{Single scattering in the dipole language. The emitted gluon with transverse coordinate $B_{\perp}$ is denoted by a double line in the large $N_{c}$ limit.}
  \label{fig:singles}
\end{figure}
The single scattering contributions are shown in \fig{fig:singles} for which the formula reads
\begin{equation}
  \label{eq:Ss}
  S(x_{\perp})=\frac{\alpha_{s} N_{c}}{2\pi^{2}}\int d^{2}B_{\perp}\frac{x_{\perp}^{2}}{B_{\perp}^{2}(B_{\perp}-x_{\perp})^{2}}\frac{d\omega}{\omega}\bigg(-\frac{\hat{q}L}{4}\bigg)\big[B_{\perp}^{2}+(B_{\perp}-x_{\perp})^{2}\big].
\end{equation}
The first part of \eq{eq:Ss} is the standard dipole probability distribution while the
\begin{gather*}
  -\frac{\hat{q}L}{4}\big[B_{\perp}^{2}+(B_{\perp}-x_{\perp})^{2}\big]
\end{gather*}
factor gives the (absorptive) single scattering amplitude for the quark-antiquark-gluon system to interact with some nucleon at impact parameter $B_{\perp}$. In the double logarithmic region where $B_{\perp}^{2}\gg x_{\perp}^{2}$, \eq{eq:Ss} becomes
\begin{equation}
  S(x_{\perp})=\bigg(-\frac{x_{\perp}^{2}}{4}\bigg)\frac{\alpha_{s} N_{c}}{\pi}\hat{q}L\int^{1/\hat{q}l_{0}}_{1/\hat{q}L}\frac{d B_{\perp}^{2}}{B_{\perp}^{2}}\int^{1/\hat{q}B_{\perp}^{4}}_{l_{0}/B_{\perp}^{2}}\frac{d \omega}{\omega}
  \label{eq:sxd}
\end{equation}
which, using \eq{eq:avgp}, leads to \eq{eq:pp}. The limits on the integrals come from the following: $\omega>l_{0}/B_{\perp}^{2}$ is just the statement that the lifetime of the fluctuation, $\omega B_{\perp}^{2}$, is greater than the nucleon size $l_{0}$. $\omega<1/\hat{q}B_{\perp}^{4}$ is the same as $1/B_{\perp}^{2}>\hat{q} B_{\perp}^{2}\omega$ as explained just below \eq{eq:tonet}. Here we are requiring $B_{\perp}$ to be small enough so that multiple scattering is not important. The lower limit on the $B_{\perp}^{2}$-integration is requiring that $B_{\perp}^{2}>x_{\perp}^{2}$ while the upper limit is necessary in order to have a nonzero range for the $\omega$-integral. The equivalence of \eq{eq:Ss} with \eq{eq:Sx} and \eq{eq:Nx} at the double logarithmic level confirms the constant factor we have taken in front of the right hand side of \eq{eq:Nx}.

\section{\label{sec:s}The single logarithmic contribution from the different boundaries}

In order to obtain a complete expression for $\avg{p_{\perp}^{2}}$, apart from the double logarithmic term, we also have to calculate the single logarithmic correction to $\avg{p_{\perp}^{2}}$, which requires to do the $dt$ and $d\omega$ integration in \eq{eq:pp} over the entire allowed phase space, i.e. $0<t<L$ and $\hat{q}l_{0}^{2}<\omega<\hat{q}L^{2}$. It is natural to divide the energy domain into two different regions, $\hat{q}l_{0}^{2}<\omega<\hat{q}l_{0}L$ and $\hat{q}l_{0}L<\omega<\hat{q}L^{2}$, and do the calculation for each region separately. The reason we make this division is that the calculation near the boundaries $(a)$ and $(c)$ are very different. Moreover, when we do the $d\omega$ and $dt$ integration over a certain region of the phase space, it is convenient to do the $t$-integral first but leave the $\omega$ variable unintegrated. The $\omega$-integration will be carried out after we combine contributions from different regions of phase space. The final expression for $\avg{p_{\perp}^{2}}$, \eq{eq:psd}, contains a double logarithmic term, as found in \eq{eq:pp}, and a single logarithmic term, which comes from crossing different boundaries $(a)$, $(b)$ and $(c)$, and an undetermined constant term. The following calculation consists of three different parts. First, we cross the boundary $(b)$ by starting from some arbitrary point $t_{0}$ in region $A$, then do the $t$-integral from $t_{0}$ to $L$, which allows us to collect the double logarithmic contribution from the region $A$ as well as the single logarithmic contribution from the boundary $(b)$. Next, we deal with the boundaries $(a)$ and $(c)$ separately. The $t$-integration for those two boundaries will be done from $0$ to $t_{0}$ such that when we add the contributions from the three regions together there is no $t_{0}$ dependence. Finally, we do the $\omega$-integration and obtain our final expression for $\avg{p_{\perp}^{2}}$. Note that crossing the boundary $(b)$ is relatively simple and in our approximation \ref{eq:Nw} is the correct formula that we should use in the regions $A$ and $B$. So in the following calculation we first calculate the contribution to $\avg{p_{\perp}^{2}}$ from regions $A$ and $B$, then extract the single logarithmic contributions from the boundaries $(a)$ and $(c)$ by modifying the dipole formula \ref{eq:Ss}. One reason we have to reformulate the dipole formula is that in \eq{eq:Nw} a time integral is used, so when crossing the boundaries $(a)$ and $(c)$ we have to express the dipole formula in terms of a time integration. After all if we want to connect calculations from different regions of phase space, we have to connect them by the same variable. The complication arises from converting the transverse coordinates integration in the dipole formula to a time integration. Another reason is that below those two boundaries certain kinematic limitation of the dipole formula will be met, the formulas no longer take the simple and compact form. We will discuss the limitations and modifications in more detail as we carry out the calculation.

\subsection{\label{sec:boundaryb}Crossing the  boundary $(b)$}

As we have seen in our previous calculation, \eq{eq:Nw} in the limit $\omega_{0}t\ll 1$ gives the same double logarithmic contribution as the single scattering contribution from the simple dipole calculation, \eq{eq:Ss}. The line $t=\sqrt{\omega/\hat{q}}$ in \fig{fig:tregion} can be taken as the separation of the single scattering and the multiple scatterings. As we start from the region $A$, crossing the boundary $(b)$, we enter the region $B$ where the multiple scatterings play the major role. In the multiple scattering region the Green function is exponentially suppressed in $B^{2}_{\perp}$ as $t$ approaches $L$, so there is no double logarithmic contribution from this region. We can choose some initial point $t_{0}$, which is in the double logarithmic region below the boundary $(b)$ but above the boundary $(a)$ or $(c)$ and satisfies $\omega_{0} t_{0}\ll 1$. Since now we are not only interested in the double logarithmic contribution from the region $A$ but also the single logarithmic contribution from the boundary $(b)$, we should use the complete expression \ref{eq:nG} in \eq{eq:Nw} to do the $t$-integration from $t_{0}$ to $L$. That is
\begin{equation}
  N(x_{\perp},\omega)=i\frac{\alpha_{s} N_{c}x_{\perp}^{2}\omega}{4\pi}\int^{L}_{t_{0}}dt\,\frac{L-t}{t^{3}}\Bigg\{\bigg(\frac{\omega_{0}t}{\sin \omega_{0}t}\bigg)^{3}\big[4-\sin^{2}\omega_{0}t\big]-4\Bigg\}.
  \label{eq:Ni}
\end{equation}
It is straightforward to do the time integration, and \eq{eq:Ni} becomes
\begin{equation}
  N(x_{\perp},\omega)=\frac{N_{c}x_{\perp}^{2}}{4\pi}\hat{q}L\bigg[\ln\frac{\omega_{0}t_{0}}{2}+\frac{1}{3}\bigg].
  \label{eq:Nb}
\end{equation}
With \eq{eq:Nb} and using \eq{eq:Sx} and \eq{eq:avgp}, we can calculate $\avg{p_{\perp}^{2}}$, but leave the $\omega$ variable unintegrated, that is
\begin{equation}
  \label{eq:boundaryb}
  \omega\frac{d}{d\omega}\avg{p_{\perp}^{2}}\bigg|_{\mathrm{boundary\ b}}=-\nabla_{x_{\perp}}^{2}N(x_{\perp},\omega)=\frac{\alpha_{s} N_{c}}{\pi}\hat{q}L\Bigg[\ln\bigg(\frac{2}{t_{0}}\sqrt{\frac{\omega}{\hat{q}}}\bigg)-\frac{1}{3}\Bigg]
\end{equation}
which not only contains the information about the double logarithmic contribution from the region $A$, but also the single logarithmic contribution from the boundary $(b)$.

\subsection{\label{sec:boundarya}Crossing the boundary $(a)$}

Our starting point is the dipole formula, \eq{eq:Ss}, but with some modifications. Near the boundary $(a)$ the transverse size of the gluon is comparable to that of the dipole. Note that in \eq{eq:Ss} a singularity occurs if we take $|B_{\perp}|\rightarrow |x_{\perp}|$. This is because we did not include the usual virtual dipole scattering term, guaranteeing probability conservation, in \eq{eq:Ss}. Indeed, the virtual term does not give a double logarithmic contribution and could be neglected in arriving at \eq{eq:sxd}. However, the virtual term is important here and can be put in simply by replacing the $-\frac{1}{4}\hat{q}L\big[B_{\perp}^{2}+(B_{\perp}-x_{\perp})^{2}\big]$ term in \eq{eq:Ss} by
\begin{equation}
  -\frac{\hat{q}L}{4}\big[B_{\perp}^{2}+(B_{\perp}-x_{\perp})^{2}-x_{\perp}^{2}\big]=-\frac{\hat{q}L}{2}B_{\perp}\cdot (B_{\perp}-x_{\perp}),
  \label{eq:news}
\end{equation}
a result which also follows directly from \eq{eq:Nx}. Moreover, we have to convert \eq{eq:Ss} to an expression with an integral over time so that the same $t_{0}$ appears in order to match onto \eq{eq:boundaryb}. Note that $x_{\perp}^{2}/B_{\perp}^{2}(B_{\perp}-x_{\perp})^{2}$ is the gluon emission probability and is calculated from squaring the wave function that a dipole emits a gluon. So we can keep the time integration in the wave function and write the gluon emission probability as
\begin{align}
  \frac{x_{\perp}^{2}}{B_{\perp}^{2}(B_{\perp}-x_{\perp})^{2}}&=-\frac{\omega^{2}}{4}\int^{0}_{-\infty}\frac{dt_{1}}{t_{1}^{2}}\int^{\infty}_{0}\frac{dt_{2}}{t_{2}^{2}}\Big[e^{-iB_{\perp}^{2}\omega/2t_{1}}B_{\perp}-e^{-i(B_{\perp}-x_{\perp})^{2}\omega/2t_{1}}(B_{\perp}-x_{\perp})\Big]\nonumber\\
  &\phantom{========}\cdot\Big[e^{i\omega B^{2}_{\perp}/2t_{2}}B_{\perp}-e^{i\omega(B_{\perp}-x_{\perp})^{2}/2t_{2}}(B_{\perp}-x_{\perp})\Big]
  \label{eq:dipolet}
\end{align}
where $t_{1}$ is the time that the gluon is emitted from the dipole and $t_{2}$ the time that the gluon is absorbed by the dipole. Substitute the modifications, \eq{eq:news} and \eq{eq:dipolet}, into \eq{eq:Ss}, do the $d^{2}B_{\perp}$ integration first, then introduce $t=t_{2}-t_{1}$ and do the $t_{1}$ integration from $0$ to $-t$. The result is
\begin{equation}
  \omega\frac{d}{d\omega}S(x_{\perp})=-\frac{\alpha_{s} N_{c}}{12\pi}\hat{q}L\,\mathrm{Re}\int^{t_{0}}_{0}dt \Bigg[\bigg(\frac{x_{\perp}^{2}}{t}+\frac{i\omega (x_{\perp}^{2})^{2}}{2t^{2}}\bigg)e^{i\omega x_{\perp}^{2}/2t}+\frac{4i}{\omega}\big(1-e^{i\omega x_{\perp}^{2}/2t}\big)\Bigg].
  \label{eq:dS}
\end{equation}
The integral in \eq{eq:dS} goes up to $t_{0}$ in order to match onto \eq{eq:boundaryb}. It is straightforward to do the $t$-integration and then take the limit where $t_{0}$ is relatively large, one finds
\begin{equation}
  \omega\frac{d}{\omega} S(x_{\perp})=\frac{\alpha_{s} N_{c} x_{\perp}^{2}}{4\pi}\hat{q}L\bigg[\ln u_{0}+\gamma-\frac{1}{3}\bigg]
  \label{eq:Sa}
\end{equation}
where $u_{0}=-i\omega x_{\perp}^{2}/2t_{0}$ and $\gamma$ is the Euler constant. The detail of the calculation is shown in Appendix~\ref{app:aintegral}. Substituting \eq{eq:Sa} in \eq{eq:avgp}, we obtain
\begin{equation}
  \omega\frac{d}{d\omega}\avg{p_{\perp}^{2}}\bigg|_{\mathrm{boundary\ a}}=\frac{\alpha_{s} N_{c}}{\pi}\hat{q}L\bigg[\ln\frac{2t_{0}}{\omega x_{\perp}^{2}}+\frac{1}{3}-\gamma\bigg].
  \label{eq:boundarya}
\end{equation}

Following the prescription in Refs.~\cite{bdmpspt,Arnold} and the discussion in \sec{sec:intro} we have taken the $\nabla_{x_{\perp}}^{2}$ in \eq{eq:avgp} to apply only to the $x_{\perp}^{2}$ prefactor on the right hand side of \eq{eq:Sa} and not to the $\ln u_{0}$ term. With this prescription, and after taking an appropriate scale for $x_{\perp}^{2}$ instead of setting $x_{\perp}^{2}=0$, $\avg{p_{\perp}^{2}}$ in \eq{eq:boundarya} corresponds to a typical $p_{\perp}^{2}$ and not the mean $p_{\perp}^{2}$ which is not a useful concept in the present context.

\subsection{\label{sec:boundaryc}Crossing boundary $(c)$}

\begin{figure}[h]
  \centering
  \subfigure[]{\label{fig:gluona}
    \includegraphics[width=3.5cm]{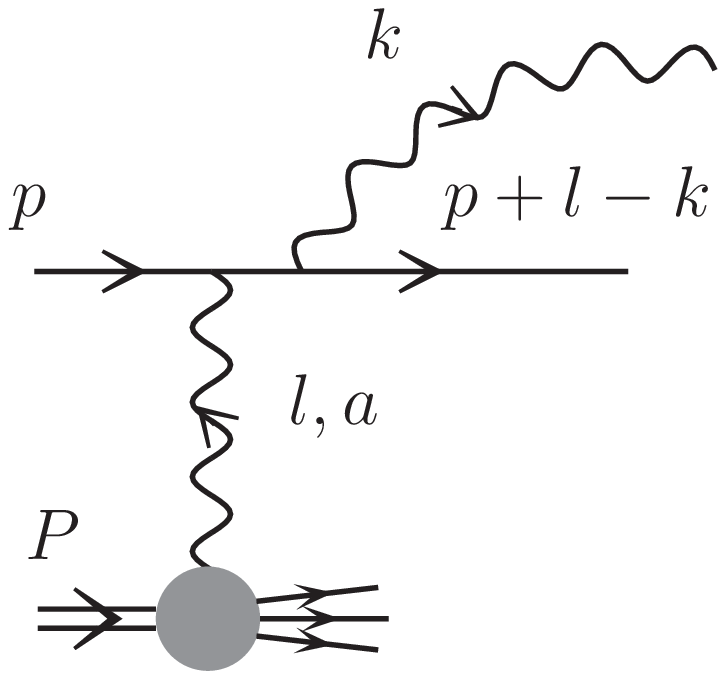}}
  \hspace{0.3in}
  \subfigure[]{\includegraphics[width=3.5cm]{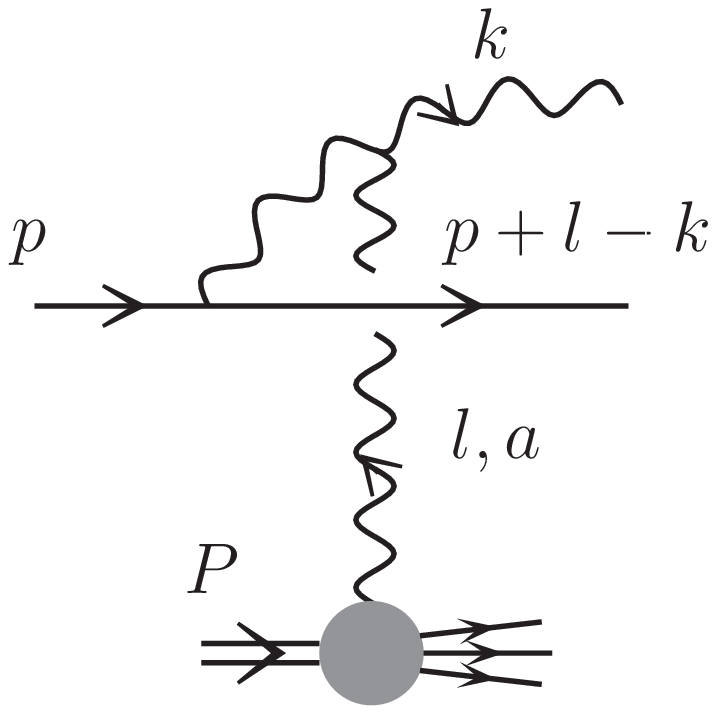}}
  \hspace{0.3in}
  \subfigure[]{\includegraphics[width=3.5cm]{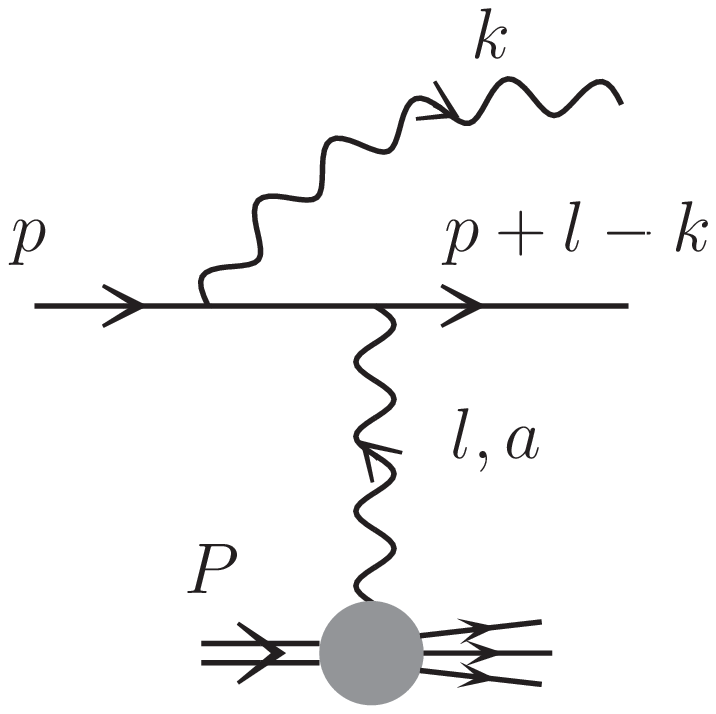}}
  \caption{Induced gluon radiation of a high-energy quark by a single scattering.}
  \label{fig:gluonemission}
\end{figure}

In order to see the limitation of \eq{eq:Ss} near the boundary $(c)$ let us first determine the $x$-value of the nucleon gluon distribution probed by the high-energy quark with a soft gluon radiation. Suppose the nucleon is at rest, $P_{\mu}=(m,0,0,0)$ where $m$ is the rest mass of the nucleon, and the quark is moving along the $z$-direction with a four-momentum $p_{\mu}=(p,0,0,p)$. Consider a typical scattering diagram, for example, \fig{fig:gluona}. $l_{-}$ from the nucleon is very close to the value $k_{\perp}^{2}/2k_{+}$. Then the $x$-value of the gluon distribution of the nucleon is
\begin{equation}
  x=\frac{l_{-}}{P_{-}}\approx \frac{k_{\perp}^{2}}{2k\cdot P}=\frac{k_{\perp}^{2}}{2m\omega}.
  \label{eq:xvalue}
\end{equation}
The lifetime of the gluon fluctuation is $t\approx 2\omega/k_{\perp}^{2}$, and we can roughly take $m\approx 1/l_{0}$. Using the above approximations, we see that $x\approx l_{0}/t$. That is near the boundary $(c)$ the $x$-value of the gluon distribution of the nucleon approaches $1$. (Note that the usual transport coefficient is assumed to be evaluated at a small value of $x$.) This means that the transport coefficient $\hat{q}$, depending on the gluon distribution can no longer be taken as a constant and its $x$-dependence cannot be neglected near boundary $(c)$.

Here we evaluate $\avg{p_{\perp}^{2}}$ of the quark by identifying it with the average $k_{\perp}^{2}$ of the radiated gluon. It is sufficient to take only the real diagrams into account, as shown in \fig{fig:gluonemission}. One can show that the virtual diagrams do not contribute to $\avg{p_{\perp}^{2}}$ in the approximation $|l_{\perp}|> |k_{\perp}|$. It is convenient to do the calculation in momentum space. Therefore from \fig{fig:gluonemission} we have
\begin{equation}
  \omega\frac{d N}{d\omega d^{2}k_{\perp}}=\frac{\alpha_{s} N_{c}}{\pi^{2}}\rho L \int d^{2}l_{\perp}\,\frac{l_{\perp}^{2}}{k_{\perp}^{2}(k_{\perp}-l_{\perp})^{2}}\frac{d \sigma}{d^{2} l_{\perp}}
\end{equation}
where $\sigma$ is the high-energy quark-nucleon scattering cross section coming from the elastic scattering and the factor $(\alpha_{s} N_{c}/\pi^{2})l_{\perp}^{2}/k_{\perp}^{2}(k_{\perp}-l_{\perp})^{2}$ is the probability that a high-energy quark emits a soft gluon. From \eq{eq:pdef} we have
\begin{equation}
  \avg{p_{\perp}^{2}}\bigg|_{\mathrm{boundary\ c}}=\int k_{\perp}^{2}\frac{d N}{d^{2}k_{\perp}}d^{2}k_{\perp}=\frac{\alpha_{s} N_{c}}{\pi^{2}}\rho L\int\frac{d \omega}{\omega}d^{2}l_{\perp}\frac{d \sigma}{d^{2}l_{\perp}}\int d^{2}k_{\perp}\frac{k_{\perp}^{2}l_{\perp}^{2}}{k_{\perp}^{2}(k_{\perp}-l_{\perp})^{2}}.
  \label{eq:pgluon}
\end{equation}
Let us write the scattering probability in terms of time integrals in order to separate the $k_{\perp}$ and $l_{\perp}$ dependence. Later the $k_{\perp}$-integration will be converted into a $x$-integration, and the $l_{\perp}$-integral will give the gluon distribution of the nucleon. So similar to \eq{eq:dipolet} we write
\begin{align}
  \int d^{2}k_{\perp}\frac{k_{\perp}^{2}l_{\perp}^{2}}{k_{\perp}^{2}(k_{\perp}-l_{\perp})^{2}}&=-\frac{1}{4\omega^{2}}\int^{0}_{-\infty}dt_{1}\int^{\infty}_{0}dt_{2}\ k_{\perp}^{2}d^{2}k_{\perp}\big[k_{\perp}e^{ik_{\perp}^{2} t_{1}/2\omega}-(k_{\perp}-l_{\perp})e^{i(k_{\perp}-l_{\perp})^{2}t_{1}/2\omega}\big]\nonumber\\
  &\phantom{====}\cdot \big[k_{\perp}e^{-ik_{\perp}^{2}t_{2}/2\omega}-(k_{\perp}-l_{\perp})e^{-i(k_{\perp}-l_{\perp})^{2}t_{2}/2\omega}\big].
  \label{eq:dipolereal}
\end{align}
Introduce $t=t_{2}-t_{1}$ and do the $t_{1}$-integral from $-t$ to $0$, \eq{eq:dipolereal} becomes
\begin{equation}
  l_{\perp}^{2}\int\frac{d^{2}k_{\perp}}{k_{\perp}^{2}}\big(1-e^{-ik_{\perp}^{2}t_{0}/2\omega}\big)
  \label{eq:kx}
\end{equation}
where we have dropped $e^{ik_{\perp}^{2}t_{0}/2\omega}$ factors not enhanced by small-$k_{\perp}^{2}$ integration. The detail derivation of \eq{eq:kx} is given in Appendix~\ref{app:bintegral}. Now we see that in \eq{eq:kx} a $k_{\perp}$-integration appears. It can be converted into an integration over $x$ using \eq{eq:xvalue}. Furthermore, the $l_{\perp}^{2}$ factor can be used to obtain the gluon distribution of the nucleon in \eq{eq:pgluon} via
\begin{equation}
  \int d^{2}l_{\perp}\, l_{\perp}^{2}\frac{d\sigma}{d^{2}l_{\perp}}=\frac{4\pi^{2}\alpha_{s} C_{F}}{N_{c}^{2}-1} x\mathcal{G}
  \label{eq:xg}
\end{equation}
where $x\mathcal{G}$ contains the general $x$-dependence of the gluon distribution of the nucleon. Putting the above two steps together we obtain 
\begin{equation}
  \avg{p_{\perp}^{2}}\bigg|_{\mathrm{boundary\ c}}=\frac{\alpha_{s} N_{c}}{\pi}\rho L \int \frac{d \omega}{\omega}\int^{1}_{0}\frac{d x}{x}\big(1-e^{-imt_{0}x})\frac{4\pi^{2}\alpha_{s} C_{F}}{N_{c}^{2}-1}x\mathcal{G}.
\end{equation}

We shall separate the gluon distribution into two different parts, one part is small at small $x$ while the other one is the gluon distribution used in the transport coefficient and has the usual small-$x$ dependence. So we write 
\begin{equation}
  x\mathcal{G}=\big(x\mathcal{G}-xG)+xG
\end{equation}
where $xG$ is the usual, small-$x$, form of the gluon distribution as used in \eq{eq:sat}. Using 
\begin{equation}
  \mathrm{Re}\int^{1}_{0}\frac{dx}{x}\big(1-e^{-ixmt_{0}}\big)=\gamma+\ln(mt_{0})+\mathcal{O}\bigg(\frac{1}{mt_{0}}\bigg)
\end{equation}
with $t_{0}$ being large, we find
\begin{equation}
  \label{eq:boundaryc}
  \omega\frac{d}{d\omega}\avg{p_{\perp}^{2}}\bigg|_{\mathrm{boundary\ c}}=\frac{\alpha_{s} N_{c}}{\pi}\hat{q}L\big[\ln(m t_{0})+\gamma\big]+\frac{\alpha_{s} N_{c}}{\pi}\hat{q}L\int^{1}_{0}\frac{d x}{x}\bigg[\frac{x\mathcal{G}}{xG}-1\bigg].
\end{equation}

\section{\label{sec:f}Final expression for $\avg{p_{\perp}^{2}}$}

Now we shall gather all our results from \sec{sec:boundaryb}, \sec{sec:boundarya} and \sec{sec:boundaryc}. In the region $\hat{q}l_{0}L<\omega<\hat{q}L^{2}$ the $t$-integration encounters the boundaries $(a)$ and $(b)$, so we add \eq{eq:boundarya} and \eq{eq:boundaryb} and do the $\omega$-integration from $\hat{q}l_{0}L$ to $\hat{q}L^{2}$. In the region $\hat{q}l_{0}^{2}<\omega<\hat{q}l_{0}L$ the $t$-integration encounters the boundaries $(c)$ and $(b)$, so we add \eq{eq:boundaryc} and \eq{eq:boundaryb} and do the $\omega$-integration from $\hat{q}l_{0}^{2}$ to $\hat{q}l_{0}L$. Finally, we can combine these two results together and obtain our final expression for the radiative corrections to the transverse momentum broadening
\begin{equation}
  \avg{p_{\perp}^{2}}=\frac{\alpha_{s} N_{c}}{8\pi}\hat{q}L\ln^{2}\frac{L^{2}}{l_{0}^{2}}+\frac{\alpha_{s}N_{c}}{\pi}\hat{q}L\bigg[\ln\frac{8ml_{0}}{\ul{x}^{2}\hat{q}L}-\frac{1}{3}+\int^{1}_{0}\frac{dx}{x}\bigg(\frac{x\mathcal{G}}{xG}-1\bigg)\bigg]\ln\frac{L}{l_{0}}+C
  \label{eq:psd}
\end{equation}
where the constant $C$ cannot be determined in our current calculation. We remind the reader that $\hat{q}Lx_{\perp}^{2}/4$ in the second term should always be taken to be of order one. (See the discussion in \sec{sec:intro}.) Note that if we rescale $l_{0}$, i.e. $l_{0}\rightarrow c\cdot l_{0}$, where $c$ is some arbitrary constant, then the first two terms in \eq{eq:psd} remain the same, however, the constant $C$ is changed by an arbitrary amount. Therefore we can only determine $\avg{p_{\perp}^{2}}$ up to some arbitrary constant, but the double and single logarithmic terms in \eq{eq:psd} are exact. 

In order to do a \textit{rough estimate} of the size of the radiative correction to $p_{\perp}$-broadening of a quark in \textit{hot} matter we take $L\simeq 5\, \mathrm{fm}$, $m\simeq 300\, \mathrm{MeV}$ and $ml_{0}\simeq 1$ as well as $x\mathcal{G}/xG\simeq 1$. (For cold matter $x\mathcal{G}/xG$ would be much smaller, perhaps something like $(1-x)^{5}$.) Then with $x_{\perp}^{2}\hat{q}L=x_{\perp}^{2}Q_{s}^{2}\simeq 4$ and $\alpha_{s}\simeq 1/3$
\begin{equation}
  \avg{p_{\perp}^{2}}_{\mathrm{rad}}\simeq 0.75\, \hat{q}L,
  \label{eq:est}
\end{equation}
a rather large contribution from radiative corrections. The dominant part of \eq{eq:est} comes from the double logarithmic part and is clearly sensitive to the exact values one takes for $\alpha_{s},L$ and $l_{0}$.

\section{\label{sec:re}Resummation and running coupling}

In this section we shall briefly discuss two extensions (modifications) of our analysis. The first of these is a resummation of our calculation while the second touches a somewhat different calculation because of the way in which we shall approach running coupling effects.

\subsection{\label{sec:resum}Resummation}

\begin{figure}[h]
  \centering
  \includegraphics[width=8cm]{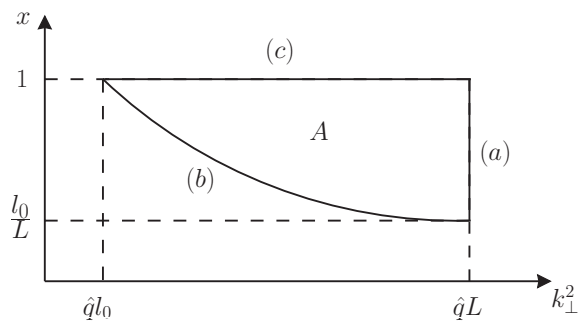}
  \caption{The double logarithmic region and its boundaries in $(x,k_{\perp}^{2})$ rather than $(t,\omega)$ of \fig{fig:tregion}.}
  \label{fig:dregion}
\end{figure}
\begin{figure}[h]
  \centering
  \subfigure[]{\label{fig:doublea}
    \includegraphics[width=5cm]{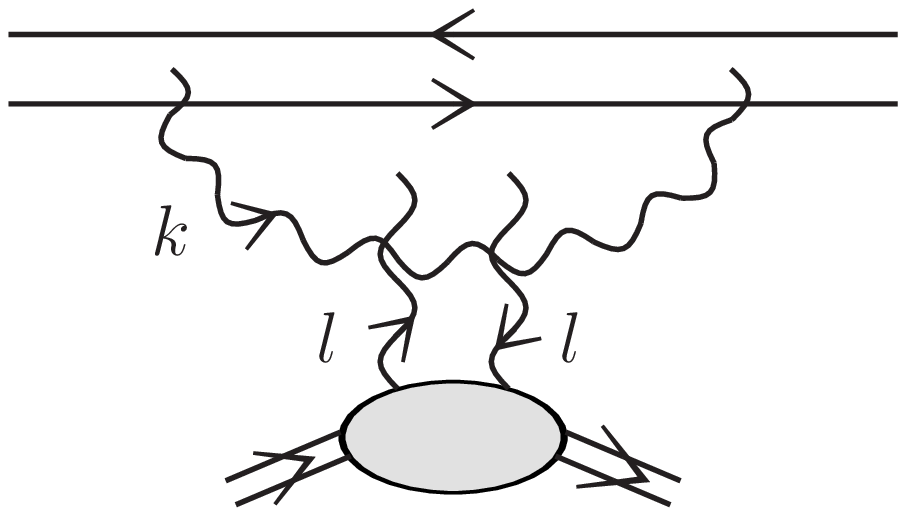}}
  \hspace{0.3in}
  \subfigure[]{\label{fig:doubleb}
    \includegraphics[width=5cm]{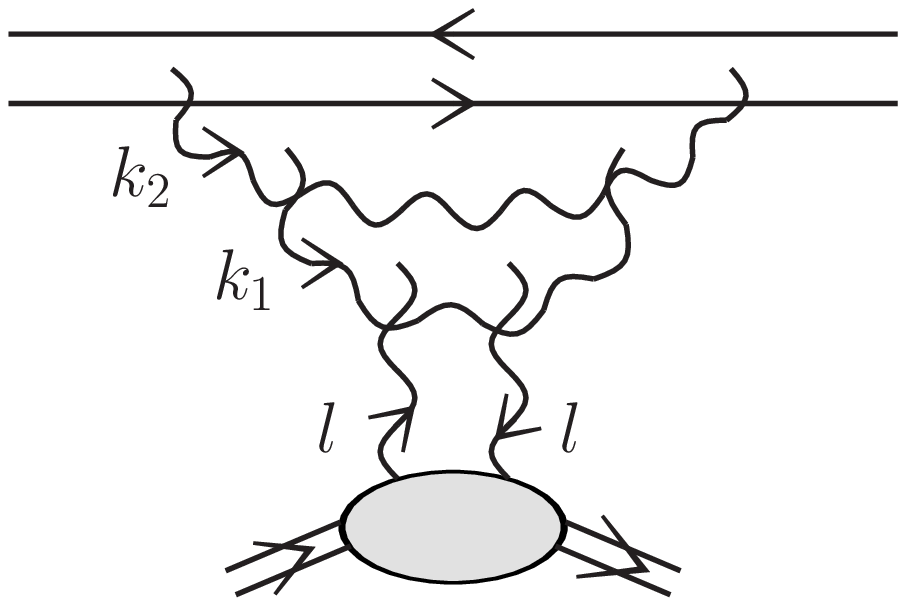}}
  \caption{Schematic illustrations of \eq{eq:repd} and \eq{eq:doubletwo} respectively.}
\end{figure}

It is not difficult to resum all leading double logarithms. To do this it is convenient to draw \fig{fig:tregion} in a somewhat different way using variables familiar from double logarithmic resummation in deep inelastic scattering, namely $k_{\perp}^{2}$ and $x$ as defined in \sec{sec:boundaryc}. The boundaries of \fig{fig:tregion} now appear as in \fig{fig:dregion}. The boundary $(b)$ now is given by
\begin{equation}
  x=\frac{\hat{q}l_{0}}{k_{\perp}^{2}}\equiv \frac{Q_{0}^{2}}{k_{\perp}^{2}}
\end{equation}
where, in the double logarithmic approximation, we do not distinguish the mass $m$ in \eq{eq:xvalue} from $l_{0}$. We shall begin by repeating the double logarithmic calculation given in \sec{sec:d} in terms of our current momentum variables. We may take the expression for $\avg{p_{\perp}^{2}}$ directly from \eq{eq:pgluon} which, using \eq{eq:sat} and \eq{eq:xg} with $x\mathcal{G}$ replaced by $xG$,  can be written as
\begin{equation}
  \avg{p_{\perp}^{2}}=\frac{\alpha_{s} N_{c}}{\pi}\hat{q}L\int^{Q_{s}^{2}}_{Q_{0}^{2}}\frac{d k_{\perp}^{2}}{k_{\perp}^{2}}\int^{1}_{Q_{0}^{2}/k_{\perp}^{2}}\frac{dx}{x}
  \label{eq:repd}
\end{equation}
and leads to
\begin{equation}
  \avg{p_{\perp}^{2}}=\frac{\alpha_{s} N_{c}}{4\pi}\hat{q}L\frac{1}{2!}\ln^{2}\bigg(\frac{L}{l_{0}}\bigg)^{2}
  \label{eq:doublefirst}
\end{equation}
exactly as in \eq{eq:pp}. The integration region in \eq{eq:repd} goes over the region $A$ in \fig{fig:dregion}. Graphically it is convenient to view \eq{eq:repd} as corresponding to the graph shown in \fig{fig:doublea} where $l_{\perp}^{2}/k_{\perp}^{2}$ is taken as usual in the logarithmic approximation. From the point of view of how the $l$-lines hook onto the quark-antiquark pair the quark-antiquark pair acts like a gluon because the quark-antiquark separation, $x_{\perp}$, is much less than $1/l_{\perp}$.
The next term in the leading double logarithmic series is illustrated in \fig{fig:doubleb} and is given by
\begin{equation}
  \avg{p_{\perp}^{2}}=\hat{q}L\bigg(\frac{\alpha_{s} N_{c}}{\pi}\bigg)^{2}\int^{Q_{s}^{2}}_{Q_{0}^{2}}\frac{d k_{2\perp}^{2}}{k_{2\perp}^{2}}\int^{k_{2\perp}^{2}}_{Q_{0}^{2}}\frac{dk_{1\perp}^{2}}{k_{1\perp}^{2}}\int^{1}_{Q_{0}^{2}/k_{1\perp}^{2}}\frac{dx_{1}}{x_{1}}\int^{x_{1}}_{Q_{0}^{2}/k_{2\perp}^{2}}\frac{dx_{2}}{x_{2}}=\hat{q}L\bigg(\frac{\alpha_{s} N_{c}}{4\pi}\bigg)^{2}\frac{1}{2!3!}\ln^{4}\bigg(\frac{L^{2}}{l_{0}^{2}}\bigg).
  \label{eq:doubletwo}
\end{equation}
The pattern is now clear and the full leading double logarithmic sum is given by
\begin{equation}
  \avg{p_{\perp}^{2}}=\hat{q}L \sqrt{\frac{4\pi}{\alpha_{s} N_{c}}}\frac{1}{\displaystyle \ln\frac{L^{2}}{l_{0}^{2}}}I_{1}\bigg[\sqrt{\frac{\alpha_{s} N_{c}}{\pi}}\ln\frac{L^{2}}{l_{0}^{2}}\bigg]
  \label{eq:doubleresum}
\end{equation}
where the nonradiative, $\hat{q}L$, term is included in \eq{eq:doubleresum} and $I_{1}(x)$ is the modified Bessel function. The effect of the boundary $(b)$ is seen by noting that if the ordered $x$ and $k_{\perp}^{2}$ integrals went between $l_{0}/L<x<1$ and $Q_{0}^{2}<k_{\perp}^{2}<Q_{s}^{2}$ without the restriction on the lower limits of the $x$ integrals in \eq{eq:repd} and \eq{eq:doubletwo} the answer would have been $\hat{q}L\, I_{0}[\sqrt{\alpha_{s} N_{c}/\pi}\ln(L^{2}/l_{0}^{2})]$ \cite{resum} rather than the right hand side of \eq{eq:doubleresum}. Though the resummation is interesting in its own right, for reasonable values of the parameters, say $L/l_{0}\simeq 5$, $\alpha_{s}=1/3$ the correction \ref{eq:doublefirst} is dominant and the resummation does not appear useful.

\subsection{\label{sec:running}Running coupling}

In all that we have done so far we have taken $\hat{q}$ to be scale independent and we have not included any scale in the coupling $\alpha_{s}$. This is what has traditionally been called the harmonic oscillator approximation. Naturally, the scale in the $\alpha_{s}$ in \eq{eq:psd} should be $Q_{s}^{2}$. However the scale in $\hat{q}$ in \eq{eq:psd} is not straightforward. For example to solve the region near boundary $(b)$ (see \fig{fig:tregion}) we would have to solve \eq{eq:Gc} for a scale dependent $\hat{q}$ to get a generalization of \eq{eq:green} and that appears beyond what we are able to do. We can, however, evaluate the double logarithmic contribution using a running coupling, and the result has some interesting features. To include running coupling effects at the double logarithmic level we take the $\hat{q}$ in \eq{eq:repd} and evaluate it at a scale $k_{\perp}^{2}$ inside the integral over $k_{\perp}^{2}$. Even this however is not completely unambiguous. In \eq{eq:sat}, using $Q_{s}^{2}=\hat{q}L$ we see that the scale comes in both in $\alpha_{s}$ and in $xG$. We will keep the scale dependence in $\alpha_{s}$ but drop it in $xG$. (In this context we note that the scale in $xG$ would involve a $\ln\ln (Q_{s}^{2}/\Lambda^{2})$ while $\alpha_{s}(Q_{s}^{2})$ would be given by $\alpha_{s}(Q_{s}^{2})=1/(b\ln Q_{s}^{2}/\Lambda^{2})$ so that the scale dependence in $xG$ is rather weak and does not change the result \eq{eq:ptrunning} below.) Then \eq{eq:repd} becomes
\begin{equation}
  \avg{p_{\perp}^{2}}=\frac{\alpha_{s}N_{c}}{\pi}\hat{q}\ln\frac{Q_{s}^{2}}{\Lambda^{2}}\int^{Q_{s}^{2}}_{Q_{0}^{2}}\frac{d k_{\perp}^{2}}{k_{\perp}^{2}}\frac{1}{\ln \frac{k_{\perp}^{2}}{\Lambda^{2}}}\int^{1}_{Q_{0}^{2}/k_{\perp}^{2}}\frac{dx}{x},
  \label{eq:prun}
\end{equation}
where now the $\hat{q}$ in \eq{eq:prun} is evaluated at scale $Q_{s}^{2}$. Identifying $\Lambda^{2}$ and $Q_{0}^{2}$ at the double logarithmic level one finds
\begin{equation}
  \avg{p_{\perp}^{2}}=\frac{\alpha_{s}N_{c}}{4\pi}\hat{q}L\ln^{2}\bigg(\frac{L}{l_{0}}\bigg)^{2}.
  \label{eq:ptrunning}
\end{equation}
What is rather remarkable here is that there are no $\ln\ln$ terms coming from the $k_{\perp}^{2}$-integration. The leading double logarithmic result here is simply a factor $2$ larger than in the fixed coupling case. Probably we should take this as a measure of uncertainty for our whole calculation. Higher orders in $(\alpha_{s}N_{c}/\pi)\ln^{2}(L^{2}/l_{0}^{2})$ can be evaluated but do not seem to lead to any simple result as happened in \eq{eq:doubleresum} for fixed coupling.

\section{\label{sec:der}Derivation of our formalism}

\begin{figure}[h]
  \centering
  \includegraphics[width=6cm]{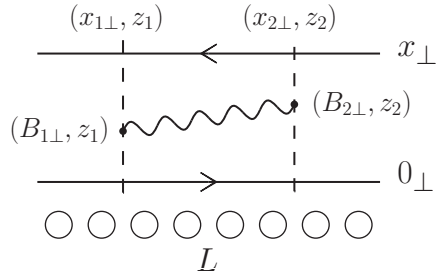}
  \caption{A quark-antiquark-gluon system in the case when the emitted gluon carries a finite fraction of the quark energy.}
  \label{fig:dipole}
\end{figure}
In this section we outline the main steps of deriving \eq{eq:Nx} in the case where the emitted gluon, real or virtual, carries a finite fraction $\xi \equiv \omega/E$ of the quark energy. More details of the derivation can be found in Appendix \ref{app:zakharov}. Let us start with the situation where the gluon is already present in the medium, as illustrated in \fig{fig:dipole}. At $z<z_1$ and $z>z_2$ the transverse coordinates of the quark and antiquark are fixed. The contribution of the multiple scatterings of the quark-antiquark pair can be written as the product of two dipoles of transverse sizes $x_{1\perp}\equiv x_\perp(z_1)$ and $x_{2\perp}\equiv x_\perp(z_2)$ multiply scattered by the medium. This dipole type multiple scattering contribution becomes the phase factor in the final expression \ref{eq:Sfull}. During the lifetime of the gluon, i.e. $z_{1}<z<z_{2}$, the quark-antiquark-gluon system is multiply scattered by the medium, and the transverse coordinate of the center of mass of the quark-gluon system is fixed relative to that of the antiquark. As a result one has
\begin{equation}
  x_{2\perp} - \xi B_{2\perp} = x_{1\perp} - \xi B_{1\perp}.\label{eq:xperp}
\end{equation}
The propagation of the quark-gluon system in the medium can be described by a Green function, $G(B_{2\perp}, z_2; B_{1\perp}, z_1)$, which can be found from the Hamiltonian\footnote{In the soft gluon and large $N_{c}$ limit the Green function defined in this section is actually the complex conjugate of that in \eq{eq:G}. However, they give the same answer since only the real part of $G$ is needed to calculate $\avg{p_{\perp}^{2}}$.}
\begin{equation}
  \label{eq:Hcb}
\hat{H} \equiv  - \frac{\nabla_{B_{2\perp}}^2}{2E \xi (1 - \xi)} - \frac{i}{4} \hat{q} \bigg\{  x_{2\perp}^2 +  \frac{N_c}{2 C_F} \big[ B_{2\perp}^2 + ( B_{2\perp} - x_{2\perp})^2  - x_{2\perp}^2  \big] \bigg\}.
\end{equation}
In fact, $G(B_{2\perp}, z_2; B_{1\perp}, z_1)$ does not depend on $x_{2\perp}$; the $x_{2\perp}$-dependence should be eliminated according to \eq{eq:xperp}. Moreover, the emission and absorption of the gluon from the quark or the antiquark can be taken into account by applying the operator
\begin{equation}
   \frac{2 g^2 C_F}{\xi(1-\xi)} P_q(\xi)\big( \nabla_{B_{1\perp}} - \xi \nabla_{x_{1\perp}} \big)\cdot\big( \nabla_{B_{2\perp}} - \xi \nabla_{x_{2\perp}} \big)
\end{equation}
to the Green function with the splitting function given by
\begin{equation}
  P_{q}(\xi) \equiv \frac{ 1 + (1-\xi)^2 }{\xi}.
\end{equation}
Putting the three contributions together, we find
\begin{align}
S(x_{2\perp})&= \frac{\alpha_s C_F}{ E^2} \mathrm{Re} \int d\xi \frac{P_{q}(\xi)}{\xi^2 (1-\xi)^2}   \int_0^L dz_2 \int_0^{z_2} dz_1\nonumber\\
&\phantom{=}\times\Big[  e^{-\frac{1}{4} \hat{q}[ x_{1\perp}^2 z_1 + x_{2\perp}^2 (L-z_2)]}  \nabla_{B_{2\perp}} \cdot \big( \nabla_{B_{1\perp}} - \xi \nabla_{x_{1\perp}} \big) G( B_{2\perp}, z_2; B_{1\perp}, z_1 )\nonumber\\
&\phantom{=}-\nabla_{B_{2\perp}} \cdot \big( \nabla_{B_{1\perp}} - \xi \nabla_{x_{1\perp}} \big) G_0( B_{2\perp}, z_2; B_{1\perp}, z_1 )\Big]\bigg|^{B_{2\perp} = x_{2\perp}, B_{1\perp} = 0, x_{1\perp} = (1-\xi) x_{2\perp} }_{B_{2\perp} = 0, B_{1\perp} = 0, x_{1\perp} = x_{2\perp} }.\label{eq:Sfull}
\end{align}
Note that $B_{1\perp}$, $B_{2\perp}$, $x_{1\perp}$ and $x_{2\perp}$ obey \eq{eq:xperp}. The $p_{\perp}$-broadening of the quark is given by
\begin{equation}
  \avg{p_\perp^2} = - \nabla_{x_{2\perp}}^2 S(x_{2})\bigg|_{x_{2\perp} = 0}.
\end{equation}
It is easy to see that in the soft gluon limit, $\xi\to 0$, \eq{eq:Sfull} reduces to \eq{eq:Nx}.

\begin{acknowledgments}
The work of Tseh Liou and A.~H.~Mueller is supported in part by the U.S. Department of Energy. Bin Wu is supported by the Agence Nationale de la Recherche project \#~11-BS04-015-01.
\end{acknowledgments}

\appendix

\section{\label{app:zakharov}Review on the path integral approach of Refs.~\cite{zlpm,zrad}}

\begin{figure}[h]
  \centering
  \subfigure[]{\label{fig:za}
    \includegraphics[width=6.5cm]{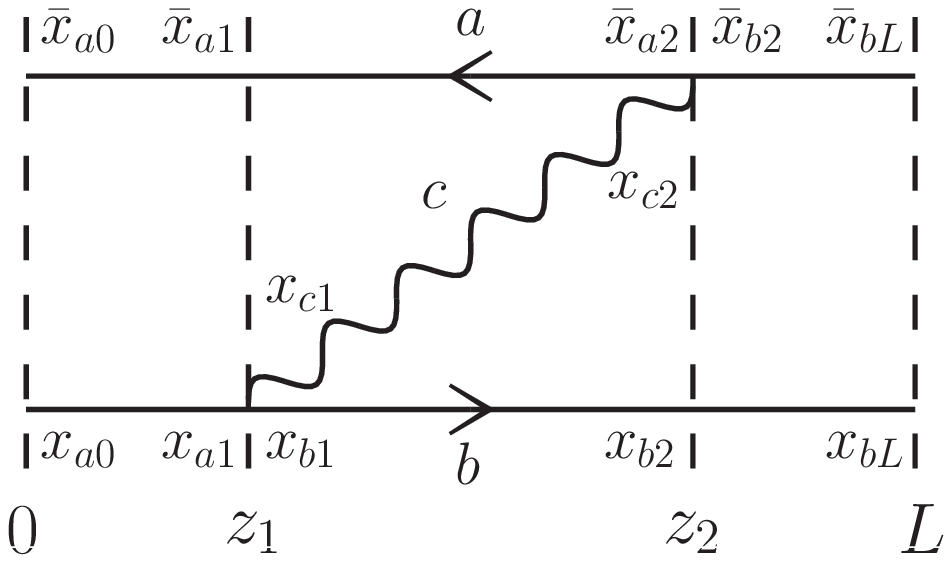}}
  \hspace{0.3in}
  \subfigure[]{\label{fig:zb}
    \includegraphics[width=6.5cm]{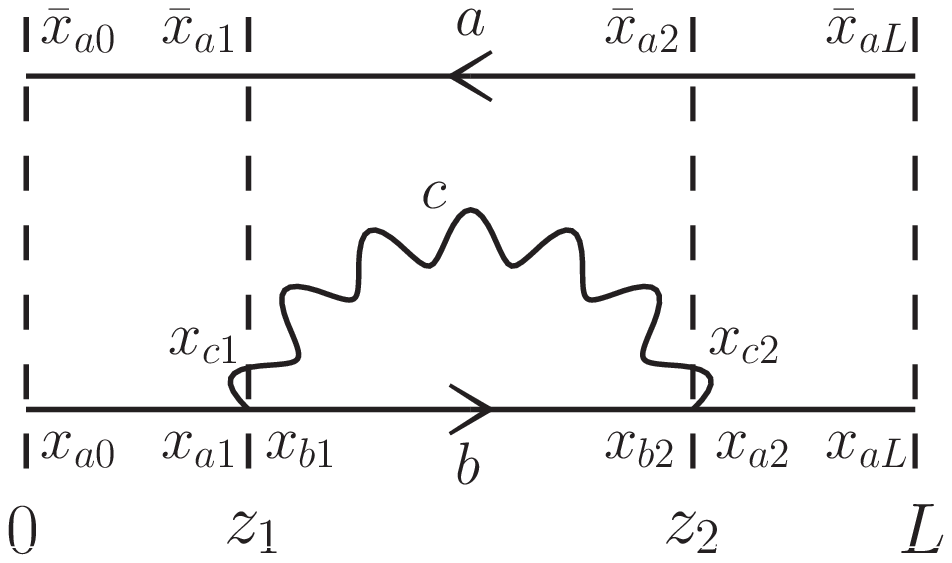}}
  \caption{Medium-induced gluon emission (\fig{fig:za}) and quark self-energy (\fig{fig:zb}). Both diagrams are composed of the multiple scattering of the quark-antiquark dipole at $z > z_2$ and $z< z_1$ and of the quark-antiquark-gluon system at $z_2>z>z_1$. During $z_2>z>z_1$, $a$, $b$ and $c$ respectively stand for the antiquark of energy $E$, the quark of energy $(1-\xi) E$ and the gluon of energy $\omega = \xi E$.}
  \label{fig:zz}
\end{figure}

In Zakharov's formalism \cite{zlpm,zrad} (see also Ref.~\cite{Blaizot} for a discussion similar to the one below), the propagator of a parton $i$ in a nuclear medium takes the form
\begin{equation}
\label{eq:propagator}
  G_i( x_{2\perp}, z_2; x_{1\perp}, z_1) =\int\limits_{x_{\perp}(z_1) = x_{1\perp},\, x_{\perp}(z_2) = x_{2\perp}} \mathcal{D}x_{\perp}\exp\bigg\{i \int_{z_1}^{z_2} dz \bigg[ \frac{E_i}{2} \dot{x}_{\perp}^2 - V_{i} ( x_{\perp}, z) \bigg]  \bigg\}
\end{equation}
where $E_i$ is the energy of the parton and $V_{i}(x_{\perp},z)$ is the potential of the medium. The momentum eigenstate of the parton in the vacuum is
\begin{equation}
  \label{eq:wavefunc}
  \psi_{i p_\perp} ( x_{\perp}, z ) \equiv \, {}_{I}\!\braket{x_{\perp},  z}{p_\perp} = e^{i p_\perp\cdot x_{\perp} - i\frac{p_\perp^2}{2 E_i} z}.
\end{equation}
We first derive explicitly the formula of $S(x)$ for real gluon emission process (\fig{fig:za}) and for  the quark self-energy (\fig{fig:zb}) inside the medium. For both processes the emitted gluon carries a finite fraction $\xi \equiv \omega/E$ of the quark energy $E$. The Feynman rules for evaluating the diagrams in \fig{fig:zz} are:
\begin{itemize}
\item At $z = 0$ and $z=L$, the wave functions of the quark and antiquark are given by \eq{eq:wavefunc} and its complex conjugate, respectively.

\item The propagators of the quark and gluon are given by \eq{eq:propagator} and the propagator of the antiquark is given by the complex conjugate of that of the quark.

\item The product of the two gluon emission/absorption vertices at $z = z_1$ and $z = z_2$, after summing over the two polarization states of the gluon, is given by
  \begin{align}
    &\int d^{2}x_{1\perp} d^{2}x_{2\perp}\, {}_I\!\bra{x_{b 1\perp},  z_1;  x_{c 1\perp},  z_1} \mathcal{H}_I(x_{1\perp}, z_1) \ket{ x_{a 1\perp},  z_1}_{I}\,{}_{I}\!\bra{ x_{a 2\perp},  z_2}\mathcal{H}_I(x_{2\perp}, z_2) \ket{ x_{b 2\perp},  z_2;  x_{c 2\perp},  z_2}_I\nonumber\\
    &=  \int \frac{d^{2}p_{1\perp}}{(2\pi)^2} \frac{d^{2}k_{1\perp}}{(2\pi)^2} \frac{d^{2}p_{2\perp}}{(2\pi)^2} \frac{d^{2}k_{2\perp}}{(2\pi)^2}\frac{2 g^2 C_F}{\xi(1-\xi)} P_q(\xi) ( k_{1\perp} - \xi p_{1\perp} )\cdot( k_{2\perp} - \xi p_{2\perp} )\nonumber\\
&\phantom{=}\times e^{i k_{1\perp}\cdot ( x_{c1\perp} - x_{b1\perp} ) - i p_{1\perp}\cdot ( x_{a1\perp} - x_{b1\perp} )-i k_{2\perp}\cdot ( x_{c2\perp} - x_{b2\perp} ) + i p_{2\perp}\cdot ( x_{a2\perp} - x_{b2\perp} )}.
  \end{align}

\item Integrate out all the transverse coordinates and the time $z_1$ and $z_2$. At the end the contribution from these two diagrams to $\frac{dN}{d^2p_\perp}$ is
  \begin{align}
    &\frac{dN_{rg}}{d^2p_\perp} \equiv \int_{0}^{1}d\xi\frac{1}{4 (2\pi)^5 \xi (1-\xi) E^2 A}\times \text{Re\big(diagram in \fig{fig:za} with $z_2 > z_1$} \big),\label{eq:Nrg}\\
&\frac{dN_{vg}}{d^2p_\perp} \equiv \int_{0}^{1}d\xi\frac{1}{4 (2\pi)^5 \xi (1-\xi) E^2 A}\times \text{Re\big(diagram in \fig{fig:zb} with $z_2 > z_1$} \big),\label{eq:Nvg}
  \end{align}
\end{itemize}
where $A$ is the area of the transverse cross section of the medium. The distribution of the transverse momentum of the quark in the final state is given by
\begin{equation}
  \frac{dN}{d^2p_\perp} = \frac{dN_{rg}}{d^2p_\perp} + \frac{dN_{vg}}{d^2p_\perp}.
\end{equation}
Eqs.~(\ref{eq:Nrg}) and (\ref{eq:Nvg}) can be simplified significantly by taking the same medium average $\avg{\cdots}$ as that in the BDMPS formalism. In this case all the path integrals can be solved analytically in the saddle-point approximation. 

The $S$-matrix for the multiple scattering of a quark-antiquark dipole is
\begin{align}
&S_i(  x_{2\perp},  x'_{2\perp}, z_2;  x_{1\perp},  x'_{1\perp}, z_1)\equiv \avg{G_i( x_{2\perp}, z_2; x_{1\perp}, z_1) G_i^*( x'_{2\perp}, z_2; x'_{1\perp}, z_1)}\nonumber\\
&=\avg{ \int \mathcal{D}x_{\perp} \mathcal{D}x'_{\perp}\, \exp\bigg\{i \int_{z_1}^{z_2} dz \bigg[ \frac{E_i}{2} {\dot{x}}^{2}_{\perp} - V_i ( x_{\perp}, z) \bigg] - i\int_{z_1}^{z_2} dz'  \bigg[ \frac{E_i}{2} \dot{x'}_{\perp}^{2} - V_i ( {x'}_{\perp}, z') \bigg]  \bigg\}}\nonumber\\
&= \int \mathcal{D}x_{\perp} \mathcal{D}{x'}_{\perp} \exp\bigg\{ \int_{z_1}^{z_2} dz \bigg[ i\frac{E_i}{2} \dot{x}_{\perp}^2 - i\frac{E_i}{2} \dot{x'}_{\perp}^2 - \frac{1}{2} \sigma_{i\,\bar{i}} ( x_{\perp} - x'_{\perp}, z ) n(z) \bigg]  \bigg\},
\label{eq:paths}
\end{align}
where
\begin{equation}
  \sigma_{i\,\bar{i}} ( x_{\perp} - x'_{\perp}, z ) n(z) \equiv \avg{\big[ V_i ( x_{\perp}, z) - V_i ( x'_{\perp},z) \big]^{2}} \simeq \frac{1}{2} \hat{q} |x_{\perp} - x'_{\perp}|^{2}.
\end{equation}
The boundary conditions $x'_{\perp}(z_{1})=x'_{1\perp}$, $x'_{\perp}(z_{2})=x'_{2\perp}$, $x_{\perp}(z_{1})=x_{1\perp}$ and $x_{\perp}(z_{2})=x_{2\perp}$ are used in \eq{eq:paths}.
It is easy to show that
\begin{align}
&\int d^2x_{1\perp} d^2 x'_{1\perp} S_i(  x_{2\perp},  x'_{2\perp}, z_2;  x_{1\perp},  x'_{1\perp}, z_1)= e^{-\frac{1}{4} \hat{q} (z_2 - z_1)| x_{2\perp} - x'_{2\perp}|^2},  \label{eq:Si}\\
&\int d^2x_{2\perp} d^2 x'_{2\perp} e^{-i p_{\perp}\cdot (x_{2\perp} - x'_{2\perp})} S_i(  x_{2\perp},  x'_{2\perp}, z_2;  x_{1\perp},  x'_{1\perp}, z_1)= e^{-i p_{\perp}\cdot (x_{1\perp} - x'_{1\perp})-\frac{1}{4} \hat{q} (z_2 - z_1) | x_{1\perp} - x'_{1\perp} |^2}.\label{eq:Sif}
\end{align}

During $z_2>z>z_1$ one needs to evaluate the following path integral
\begin{align}
  \label{eq:3body}
&G_{abc}( x_{ a 2\perp},   x_{b 2\perp},   x_{c 2\perp}, z_2; x_{ a 1\perp},   x_{b 1\perp},   x_{c 1\perp}, z_1 )\nonumber\\
& \equiv\avg{ G_b(  x_{b 2\perp},  z_2 ;   x_{b 1\perp},  z_1  ) G_c(  x_{c 2\perp},  z_2 ;   x_{c 1\perp},  z_1  ) G^*_a( x_{ a 2\perp},  z_2 ;  x_{ a 1\perp},  z_1  ) }\nonumber\\
&=\avg{\int \mathcal{D} x_{a\perp} \mathcal{D} x_{b\perp} \mathcal{D} x_{c\perp}\, \exp\bigg\{i\int dz \bigg[ \frac{E_b}{2} \dot{x}_{b\perp}^2 + \frac{E_c}{2} \dot{x}_{c\perp}^2 - \frac{E_a}{2} \dot{x}_{a\perp}^2- V_b ( x_{b\perp}, z )  - V_c ( x_{c\perp}, z ) + V_a ( x_{a\perp}, z )\bigg]\bigg\}} \nonumber\\
&=  \int \mathcal{D} x_{a\perp} \mathcal{D} x_{b\perp} \mathcal{D} x_{c\perp}\, \exp\bigg\{\int dz \bigg[ i\frac{E_b}{2} \dot{x}_{b\perp}^2 + i\frac{E_c}{2} \dot{x}_{c\perp}^2 - i\frac{E_a}{2} \dot{x}_{a\perp}^2 - \frac{1}{2} \sigma_{abc}( x_{ab\perp}, x_{cb\perp} ) n(z) \bigg]\bigg\}
\end{align}
where
\begin{align}
  &\frac{1}{2}\sigma_{abc}( x_{ab\perp}, x_{cb\perp} ) n(z)\equiv \avg{ | V_b ( x_{b\perp}, z )  + V_c ( x_{c\perp}, z ) - V_a ( x_{a\perp}, z ) |^2} \nonumber\\
&\simeq \frac{1}{4}\hat{q} \bigg\{  x_{ab\perp}^2 +  \frac{N_c}{2 C_F} \big[ x_{cb\perp}^2 + ( x_{cb\perp} - x_{ab\perp})^2  - x_{ab\perp}^2  \big] \bigg\}
\end{align}
with
\begin{equation}
  x_{ab\perp}\equiv x_{a\perp} - x_{b\perp},\quad x_{cb\perp}\equiv x_{c\perp} - x_{b\perp}. 
\end{equation}
Here indices $a$, $b$ and $c$ stand for the antiquark of energy $E$, the quark of energy $(1-\xi) E$ and the gluon of energy $\omega = \xi E$, respectively. The boundary conditions in the definition of $G_{abc}$ in \eq{eq:3body} are $x_{a\perp}(z_{1})=x_{a1\perp}$, $x_{a\perp}(z_{2})=x_{a2\perp}$, $x_{b\perp}(z_{1})=x_{b1\perp}$, $x_{b\perp}(z_{2})=x_{b2\perp}$, $x_{c\perp}(z_{1})=x_{c1\perp}$ and $x_{c\perp}(z_{2})=x_{c2\perp}$. The evaluation of the path integral in \eq{eq:3body} amounts to calculating the classical trajectory of the partons with the Lagrangian
\begin{equation}
  L = \frac{E}{2} \xi (1 - \xi) \dot{x}_{cb\perp}^2 + E ( \xi \dot{x}_{cb\perp} - \dot{x}_{ab\perp} )\cdot \dot{\bar{x}}_{\perp} + \frac{i}{2} \sigma_{abc}( x_{ab\perp}, x_{cb\perp} ) n(z),
  \label{eq:L}
\end{equation}
where
\begin{equation}
  \bar{x}_{\perp} \equiv \frac{1}{2}\big[ (1 - \xi) x_{b\perp} + \xi x_{c\perp} + x_{a\perp} \big].
\end{equation}
As a result one finds \cite{ZFormalism}
\begin{align}
  \label{eq:Gabc}
G_{abc} &= G^0_a\big( \xi x_{c2\perp} + (1 - \xi) x_{b2\perp}, z_2; \xi x_{c1\perp} + (1 - \xi) x_{b1\perp}, z_1  \big)\nonumber\\
&\phantom{=}\times G^{0*}_a( x_{a2\perp}, z_2; x_{a1\perp}, z_1  )  G(x_{cb2\perp}, z_2; x_{cb1\perp}, z_1),
\end{align}
where the propagator $G$ is defined by the Hamiltonian 
\begin{align}
  H_{bc} &\equiv - \frac{\nabla_{x_{cb\perp}}^2}{2E \xi (1 - \xi)} - \frac{i}{2} \sigma_{abc}( x_{ab\perp}, x_{cb\perp}) n(z),\nonumber\\
&\simeq - \frac{\nabla_{x_{cb\perp}}^2}{2E\xi (1 - \xi)} - \frac{i}{4} \hat{q} \bigg\{  x_{ab\perp}^2 +  \frac{N_c}{2 C_F} \big[ x_{cb\perp}^2 + ( x_{cb\perp} - x_{ab\perp})^2  - x_{ab\perp}^2  \big]\bigg\}
\label{eq:hfull}
\end{align}
with
\begin{align}
&x_{ab\perp}(z) - \xi x_{cb\perp}(z)\nonumber\\
&= x_{ab\perp}(z_1) - \xi x_{cb\perp}(z_1) + \frac{z - z_1}{z_2 - z_1}\big[ x_{ab\perp}(z_2) - \xi x_{cb\perp}(z_2) - x_{ab\perp}(z_1) + \xi x_{cb\perp}(z_1) \big]\nonumber\\
&\equiv x_{ab1\perp} - \xi x_{1cb\perp} + \frac{z - z_1}{z_2 - z_1}\big[ x_{ab2\perp} - \xi x_{cb2\perp} - x_{ab1\perp} + \xi x_{cb1\perp} \big],
  \label{eq:xabz}
\end{align}
and
\begin{equation}
  G_i^0( x_{\perp}, z; 0 , 0) = \frac{E_i}{ 2 \pi i z} e^{\frac{i E_i x_{\perp}^2}{2z}}\theta(z).
\end{equation}
Insert (\ref{eq:Si}), (\ref{eq:Sif}) and (\ref{eq:Gabc}) into (\ref{eq:Nrg}) and (\ref{eq:Nvg}), and after some algebra we have
\begin{align}
  \label{eq:Ix}
S_{rg}(x_{2\perp}) &\equiv \int d^2p_\perp\, e^{-i p_\perp\cdot x_{2\perp}} \frac{dN_{rg}}{d^2p_\perp}\nonumber\\
&= \frac{\alpha_s C_F}{ E^2} \mathrm{Re} \int d\xi \frac{P_{q}(\xi)}{\xi^2 (1-\xi)^2}   \int_0^L dz_2 \int_0^{z_2} dz_1 e^{-\frac{1}{4} \hat{q}[ (1-\xi)^2 z_1 + (L-z_2) ] x_{2\perp}^2} \nonumber\\
&\phantom{=}\times \nabla_{ x_{cb2\perp}} \cdot  (\nabla_{  x_{cb1\perp}} + \xi \nabla_{  x_{ab1\perp}}) G( x_{cb2\perp}, z_2; x_{cb1\perp}, z_1 )\bigg|_{x_{cb1\perp} = 0,\ x_{ab1\perp} = (1-\xi) x_{2\perp},\ x_{cb2\perp} = x_{2\perp}},\\
S_{vg}(x_{2\perp}) &\equiv  \int d^2p_\perp\, e^{-i p_\perp\cdot x_{2\perp}} \frac{dN_{vg}}{d^2p_\perp}\nonumber\\
&= -\frac{\alpha_s C_F}{ E^2} \mathrm{Re}\int d\xi \frac{P_{q}(\xi)}{\xi^2 (1-\xi)^2}   \int_0^L dz_2 \int_0^{z_2} dz_1 e^{-\frac{1}{4} \hat{q} ( L - z ) x_{2\perp}^2} \nonumber\\
&\phantom{=}\times\nabla_{ x_{cb2\perp}} \cdot  (\nabla_{  x_{cb1\perp}} + \xi \nabla_{  x_{ab1\perp}})  G( x_{cb2\perp}, z_2; x_{cb1\perp}, z_1 )\bigg|_{x_{cb2\perp} =0= x_{cb1\perp},\ x_{ab1\perp} = x_{ab2\perp} = x_{2\perp} },
\end{align}
and $S(x_{2\perp})=S_{rg}(x_{2\perp})+S_{vg}(x_{2\perp})$. Here $G$ is the propagator defined by $H_{bc}$ in \eq{eq:hfull} with 
\begin{equation}
  x_{ab\perp}(z)= x_{ab1\perp} +\xi\big(x_{cb\perp}(z)- x_{cb1\perp}\big)
\end{equation}
and as a result $G$ only depends on $x_{ab1\perp}$, $x_{cb1\perp}$ and $x_{cb2\perp}$. \eq{eq:Sfull} can be obtained by replacing $x_{ab\perp}$ by $x_\perp$ and $x_{cb\perp}$ by $B_\perp$ and subtracting out the limit $\hat{q}\to 0$.

\section{Two integrals}
\subsection{\label{app:aintegral}First one}

In this appendix we evaluate the integral
\begin{equation}
  I_{1}=\int d^{2}B_{\perp}\, B_{\perp}\cdot (B_{\perp}-x_{\perp})\frac{x_{\perp}^{2}}{B_{\perp}^{2}(B_{\perp}-x_{\perp})^2}.
\end{equation}
With the help of \eq{eq:dipolet} $I_{1}$ can be written as
\begin{align}
  &I_{1}=-\frac{\omega^{2}}{4}\int^{0}_{-\infty}\frac{d t_{1}}{t_{1}^{2}}\int^{\infty}_{0}\frac{d t_{2}}{t_{2}^{2}}\int d^{2}B_{\perp}\, B_{\perp}\cdot (B_{\perp}-x_{\perp})\Bigg\{B_{\perp}^{2}\exp\bigg[-\frac{1}{2}iB_{\perp}^{2}\omega\Big(\frac{1}{t_{1}}-\frac{1}{t_{2}}\Big)\bigg]\nonumber\\
  &-B_{\perp}\cdot (B_{\perp}-x_{\perp})\exp\bigg[-i\frac{B_{\perp}^{2}\omega}{2t_{1}}+i\frac{(B_{\perp}-x_{\perp})^{2}\omega}{2t_{2}}\bigg]-B_{\perp}\cdot (B_{\perp}-x_{\perp})\exp\bigg[i\frac{B_{\perp}^{2}\omega}{2t_{2}}-i\frac{(B_{\perp}-x_{\perp})^{2}\omega}{2t_{1}}\bigg]\nonumber\\
  &+(B_{\perp}-x_{\perp})^{2}\exp\bigg[i\frac{1}{2}(B_{\perp}^{2}-x_{\perp})^{2}\omega\Big(\frac{1}{t_{2}}-\frac{1}{t_{1}}\Big)\bigg]\Bigg\}.
  \label{eq:I}
\end{align}
The first term and the last term in \eq{eq:I} give the same result after the integration. It is straightforward to do the $B_{\perp}$-integration, the first term in \eq{eq:I} gives
\begin{equation}
  \label{eq:first}
  -\frac{4\pi i}{\omega}\int^{0}_{-\infty}dt_{1}\int^{\infty}_{0}dt_{2}\frac{t_{1}t_{2}}{(t_{2}-t_{1})^{3}}=-\frac{4\pi i}{\omega}\int^{t_{0}}_{0}\frac{dt}{t^{3}}\int^{0}_{-t}dt_{1}\, t_{1}(t_{1}+t)=\frac{\pi x_{\perp}^{2}}{3u_{0}}
\end{equation}
where $u_{0}=-i\omega x_{\perp}^{2}/2t_{0}$. The second and third term in \eq{eq:I} give the same result. Again, one can do the $B_{\perp}$-integration first and find
\begin{equation}
  \label{eq:thirdone}
  \frac{\pi \omega}{2}\int^{0}_{-\infty}dt_{1}\int^{\infty}_{0}dt_{2}\frac{e^{i\omega x_{\perp}^{2}/[2(t_{2}-t_{1})]}}{(t_{2}-t_{1})^{2}}\Bigg[i\frac{8t_{1}t_{2}}{\omega^{2}(t_{2}-t_{1})}-\frac{x_{\perp}^{2}}{\omega}-\frac{8t_{1}t_{2}}{\omega(t_{2}-t_{1})^{2}}x_{\perp}^{2}-i\frac{t_{1}t_{2}}{(t_{2}-t_{1})^{3}}x_{\perp}^{4}\Bigg].
\end{equation}
Introduce the time difference $t=t_{2}-t_{1}$ and do the $t_{1}$-integral from $-t$ to $0$ as in \eq{eq:first}, then \eq{eq:thirdone} becomes
\begin{equation}
  \label{eq:thirdtwo}
  \frac{\pi}{3}\int^{t_{0}}_{0}dt\, e^{i\omega x_{\perp}^{2}/2t}\bigg[-\frac{2i}{\omega}+\frac{x_{\perp}^{2}}{2t}+\frac{i\omega x_{\perp}^{4}}{4t^{2}}\bigg].
\end{equation}

With the help of the incomplete gamma function $\Gamma(\alpha,x)$, we can write the three $t$-integrals as
\begin{equation*}
  \int^{t_{0}}_{0}dt\, e^{i\omega x_{\perp}^{2}/2t}=-\frac{i\omega x_{\perp}^{2}}{2}\Gamma(-1,u_{0}),\ \int^{t_{0}}_{0}dt \, \frac{e^{i\omega x_{\perp}^{2}/2t}}{t}=\Gamma(0,u_{0}),\ \int_{0}^{t_{0}}dt \frac{e^{i\omega x_{\perp}^{2}/2t}}{t^{2}}=\frac{2i}{\omega x_{\perp}^{2}}\Gamma(1,u_{0}).
\end{equation*}
Then \eq{eq:thirdtwo} becomes
\begin{equation}
  \label{eq:second}
  \frac{\pi}{3}x_{\perp}^{2}\bigg[-\Gamma(-1,u_{0})+\frac{1}{2}\Gamma(0,u_{0})-\frac{1}{2}\Gamma(1,u_{0})\bigg].
\end{equation}
Finally, combining \eq{eq:second} and \eq{eq:first} we arrive at
\begin{equation}
  \label{eq:If}
  I_{1}=\frac{2\pi x_{\perp}^{2}}{3}\bigg[\frac{1}{u_{0}}-\Gamma(-1,u_{0})+\frac{1}{2}\Gamma(0,u_{0})-\frac{1}{2}\Gamma(1,u_{0})\bigg].
\end{equation}
Using the properties of the incomplete gamma function $\Gamma(\alpha+1,x)=\alpha\Gamma(\alpha,x)+x^{\alpha}e^{-x}$ and the expansion $\Gamma(0,x)=-\gamma-\ln x$, for $x\ll 1$, we can further simplify \eq{eq:If} in the limit $u_{0}\rightarrow 0$ as
\begin{equation}
  I_{1}=-\pi x_{\perp}^{2}\bigg[\ln u_{0}+\gamma-\frac{1}{3}\bigg].
\end{equation}

\subsection{\label{app:bintegral}Second one}
Here we evaluate the following integral
\begin{gather*}
  I_{2}=\int d^{2}k_{\perp}\, \frac{k^{2}_{\perp}l_{\perp}^{2}}{k_{\perp}^{2}(k_{\perp}-l_{\perp})^{2}}.
\end{gather*}
Again we rewrite it in terms of time integrals as in \eq{eq:dipolereal}, and find
\begin{align}
  I_{2}&=-\frac{1}{4\omega^{2}}\int^{0}_{-\infty}dt_{1}\int^{\infty}_{0}dt_{2}\int d^{2}k_{\perp}\, k_{\perp}^{2}\Bigg\{k_{\perp}^{2}e^{ik_{\perp}^{2}(t_{1}-t_{2})/2\omega}+(k_{\perp}-l_{\perp})^{2}e^{i(k_{\perp}-l_{\perp})^{2}(t_{1}-t_{2})/2\omega}\nonumber\\
  &\phantom{==}-k_{\perp}\cdot (k_{\perp}-l_{\perp})\exp\bigg[i\frac{(t_{1}-t_{2})}{2\omega}\Big(k_{\perp}+\frac{t_{2}l_{\perp}}{t_{1}-t_{2}}\Big)^{2}-i\frac{l_{\perp}^{2}}{2\omega}\frac{t_{1}t_{2}}{t_{1}-t_{2}}\bigg]\nonumber\\
  &\phantom{==}-k_{\perp}\cdot (k_{\perp}-l_{\perp})\exp\bigg[i\frac{(t_{1}-t_{2})}{2\omega}\Big(k_{\perp}-\frac{t_{1}l_{\perp}}{t_{1}-t_{2}}\Big)^{2}-i\frac{l_{\perp}^{2}}{2\omega}\frac{t_{1}t_{2}}{t_{1}-t_{2}}\bigg]\Bigg\}.
  \label{eq:itwo}
\end{align}
For the third and forth terms in \eq{eq:itwo}, we shift the $k_{\perp}$ variable according to $k_{\perp}\rightarrow k_{\perp}-t_{2}l_{\perp}/(t_{1}-t_{2})$ and $k_{\perp}\rightarrow k_{\perp}+t_{1}l_{\perp}/(t_{1}-t_{2})$ respectively in order to make all the terms to have the same phase factor. Averaging the direction of $k_{\perp}$, we find
\begin{equation}
  I_{2}=-\frac{l_{\perp}^{2}}{4\omega^{2}}\int d^{2}k_{\perp}\int^{0}_{-\infty}dt_{1}\int^{\infty}_{0}dt_{2}\bigg\{\bigg[1-\frac{2(t_{1}+t_{2})^{2}}{(t_{1}-t_{2})^{2}}\bigg]k_{\perp}^{2}+\frac{i}{\omega}\frac{t_{1}t_{2}}{t_{1}-t_{2}}(k_{\perp}^{2})^{2}\bigg\}e^{-ik_{\perp}^{2}t/2\omega}.
\end{equation}
Introduce $t=t_{2}-t_{1}$, and it is straightforward to do the $t_{1}$-integration which leads to
\begin{equation}
  I_{2}=-\frac{il_{\perp}^{2}}{6\omega}\int d^{2}k_{\perp}\int^{t_{0}}_{0}dt\, t\bigg(\frac{\partial}{\partial t}-t\frac{\partial^{2}}{\partial t^{2}}\bigg)e^{-ik_{\perp}^{2}t/2\omega}=l_{\perp}^{2}\int \frac{dk_{\perp}^{2}}{k_{\perp}^{2}}\Big(1-e^{-ik_{\perp}^{2}t/2\omega}\Big)
\end{equation}
where in the last step we have neglected the $e^{-k_{\perp}^{2}t_{0}/2\omega}$ terms.


\begin{thebibliography}{99}
\bibitem{Bodwin} 
  G.~T.~Bodwin, S.~J.~Brodsky and G.~P.~Lepage,
  Phys.\ Rev.\ Lett.\  {\bf 47}, 1799 (1981).

\bibitem{bdmpspt} 
  R.~Baier, Y.~L.~Dokshitzer, A.~H.~Mueller, S.~Peigne and D.~Schiff,
  Nucl.\ Phys.\ B {\bf 484}, 265 (1997)
  [hep-ph/9608322].

\bibitem{Kovchegov} 
  Y.~V.~Kovchegov and K.~Tuchin,
  Phys.\ Rev.\ D {\bf 65}, 074026 (2002)
  [hep-ph/0111362].
  
\bibitem{Mueller} 
  A.~H.~Mueller and S.~Munier,
  Nucl.\ Phys.\ A {\bf 893}, 43 (2012)
  [arXiv:1206.1333 [hep-ph]].

\bibitem{BW} 
  B.~Wu,
  JHEP {\bf 1110}, 029 (2011)
  [arXiv:1102.0388 [hep-ph]].

\bibitem{bdmpsenergy} 
  R.~Baier, Y.~L.~Dokshitzer, A.~H.~Mueller, S.~Peigne and D.~Schiff,
  Nucl.\ Phys.\ B {\bf 483}, 291 (1997)
  [hep-ph/9607355];

\bibitem{bdmpsz} 
  R.~Baier, Y.~L.~Dokshitzer, A.~H.~Mueller and D.~Schiff,
  Nucl.\ Phys.\ B {\bf 531}, 403 (1998)
  [hep-ph/9804212].

\bibitem{zlpm} 
  B.~G.~Zakharov,
  JETP Lett.\  {\bf 63}, 952 (1996)
  [hep-ph/9607440].

\bibitem{zrad} 
  B.~G.~Zakharov,
  JETP Lett.\  {\bf 65}, 615 (1997)
  [hep-ph/9704255].

\bibitem{ZFormalism} 
  R.~Baier, D.~Schiff and B.~G.~Zakharov,
  Ann.\ Rev.\ Nucl.\ Part.\ Sci.\  {\bf 50}, 37 (2000)
  [hep-ph/0002198].

\bibitem{dipolebfkl} 
  A.~H.~Mueller,
  Nucl.\ Phys.\ B {\bf 415}, 373 (1994).

\bibitem{book}
  Y.~V.~ Kovchegov and E.~.M.~ Levin, ``Quantum Chromodynamics at High Energy'', Cambridge Univ. Press (2012).

\bibitem{Arnold} 
  P.~B.~Arnold,
  Phys.\ Rev.\ D {\bf 80}, 025004 (2009)
  [arXiv:0903.1081 [nucl-th]].

\bibitem{mv} 
  L.~D.~McLerran and R.~Venugopalan,
  Phys.\ Rev.\ D {\bf 49}, 2233 (1994)
  [hep-ph/9309289].

\bibitem{resum}
  See Sec.~2.4.6 in \cite{book}.

\bibitem{Blaizot} 
  J.~-P.~Blaizot, F.~Dominguez, E.~Iancu and Y.~Mehtar-Tani,
  JHEP {\bf 1301}, 143 (2013)
  [arXiv:1209.4585 [hep-ph]].

\end{thebibliography}
\end{document}